\documentclass[
prd,
superscriptaddress,
10 pt,
preprintnumbers,
notitlepage,
secnumarabic,
nobibnotes,
nofootinbib,
showpacs
]{revtex4-1}

\usepackage[T1]{fontenc}
\usepackage{bm}
\usepackage[pdftex]{color,graphicx}
\usepackage[makeroom]{cancel}
\usepackage{amssymb}
\usepackage{amsmath}
\usepackage{tabularx}
\usepackage{dsfont}
\usepackage{hyperref}
\usepackage{manfnt}
\usepackage{tikz}
\usetikzlibrary{shapes}
\usetikzlibrary{arrows}
\tikzset{l/.style={draw=black, line width=1pt}}
\tikzset{s/.style={auto, outer sep=-1}}
\tikzset{t/.style={auto, outer sep=2, pos=-0.1}}
\tikzset{b/.style={auto, outer sep=2, pos=1.1}}
\tikzset{i/.style={auto, outer sep=-1}}
\tikzset{j/.style={auto, outer sep=-2, pos=0.8}}
\tikzset{a/.style={pos=1.1}}

\usepackage{perpage}
\MakePerPage{footnote}

\newcommand{\bra}[1]{\ensuremath{\left\langle#1\right|}}
\newcommand{\ket}[1]{\ensuremath{\left|#1\right\rangle}}
\newcommand{\bracket}[2]{\ensuremath{\left\langle #1 \middle| #2 \right\rangle}}

\newcommand{\vket}[1]{\ensuremath{\Biggl|\vcenter{\hbox{#1}}\hspace{-5pt}\Biggr\rangle}}

\newcommand\atopp[2]{\genfrac{}{}{0pt}{}{#1}{#2}}

\begin{document}
\title{Quantum reduced loop gravity: extension to scalar field}

\author{Jakub Bilski}
\email{jakubbilski14@fudan.edu.cn}
\affiliation{Department of Physics, Fudan University, 200433 Shanghai, China.}

\author{Emanuele Alesci}
\email{Emanuele.Alesci@fuw.edu.pl}
\affiliation{Instytut Fizyki Teoretycznej, Uniwersytet Warszawski, ulica Ho\.za 69, 00-681 Warszawa, Poland.}

\author{Francesco Cianfrani} 
\email{francesco.cianfrani@ift.uni.wroc.pl}
\affiliation{Instytut Fizyki Teoretycznej, Uniwersytet Wroc\l{}awski, plac Maksa Borna 9, 50-204 Wroc\l{}aw, Poland.}

\begin{abstract}
The quantization of the Hamiltonian for a scalar field is performed in the framework of Quantum Reduced Loop Gravity. We outline how the regularization can be performed by using the analogous tools adopted in full Loop Quantum Gravity and the matrix elements of the resulting operator between basis states are analytic coefficients. These achievements open the way for a consistent analysis of the Quantum Gravity corrections to the classical dynamics of gravity in the presence of a scalar field in a cosmological setting. 
\end{abstract}
\maketitle

\section{Introduction}

Quantum reduced loop gravity (QRLG) is a framework that describes the gravitational field of systems, whose spatial part of metric and whose dreibein's are gauge-fixed to a diagonal form. It was introduced in \cite{Alesci:2012md,Alesci:2013lea} and developed in \cite{Alesci:2013xd}-\cite{Alesci:2015nja}. The theory has been successfully applied to an inhomogeneous extension of the Bianchi I model \cite{Alesci:2013xd,Alesci:2015jca}.

QRLG is constructed from Loop Quantum Gravity (LQG) \cite{Ashtekar:2004eh,Rovelli:2004tv,Thiemann:2007zz} by imposing weakly gauge-fixing conditions in the kinematical Hilbert space. Therefore, it is a direct application of LQG that describes quantum cosmology and it differs from Loop Quantum Cosmology (LQC) \cite{Bojowald:2011zzb,Ashtekar:2011ni,Banerjee:2011qu}, in which quantization is performed in minisuperspace, {\it i.e.} after reducing the phase space on a classical level. The semiclassical limit of QRLG reproduces the quantum constraint adopted in LQC \cite{Alesci:2014uha} and it also leads to enhanced inverse volume corrections \cite{Alesci:2014rra,Alesci:2015nja}. 

The most significant implication of LQC and QRLG is that the initial singularity is replaced by the Big Bounce \cite{Ashtekar:2006rx}. The only possibility to test such a prediction is via the modifications it implies in the scalar spectrum of perturbations \cite{Bojowald:2011hd,Agullo:2012sh,Calcagni:2012vw}. These modifications are due to the quantum corrections to the dynamics of the gravitational and inflaton fields. However, till now QRLG has been realized only in vacuum. Vacuum solutions in cosmology do not have real physical meaning for two reasons. First, Universe is filled with matter, which plays a peculiar role during its evolution. For instance, the photons of cosmic microwave background radiation provides the best source of information on its structure, while an inflationary phase can be realized through a slow-rolling scalar field. Second, time in General Relativity (GR) is not an observable and a natural way to account for it is via the introduction of a clock-matter field.


In this paper, we introduce a scalar field in QRLG and we define the operator corresponding to its contribution to the scalar constraint. We quantize the field according with the LQG procedure given in \cite{Thiemann:1997rt,Thiemann:1997rq}. Basic quantum variables are point holonomies and smeared momenta leading to polymer representation 
\cite{Ashtekar:2002sn,Ashtekar:2002vh,Kaminski:2005nc,Kaminski:2006ta}. The associated scalar constraint is quantized via a regularization of the classical expression, which provides a constraint written entirely in terms of $SU(2)$ and point holonomies, together with the corresponding smeared momenta. This formulation is completely under control technically. This formulation is here adapted to QRLG, where the volume operator, thus all the relevant computations, are analytic. The final outcome of our analysis are precisely the analytic matrix elements of the scalar part of the scalar constraint between the basis elements of QRLG, which is the starting point for future applications.

In particular, we give an introduction to QRLG in section \ref{II} by defining all the relevant structures of the kinematical Hilbert space. We focus our attention on states based at graphs having six-valent nodes \cite{Alesci:2015nja}, which allow us to construct a cubulation of the whole spatial manifold. In section \ref{III}, the classical and quantum formulation for a scalar field is given. On a classical level, we write the contributions of the scalar field to the scalar and vector constraints. The regularization of the field contribution to the scalar constraint is performed in section \ref{IV}. We just adapt the regularization performed in \cite{Thiemann:1997rt} to our case, which means replacing the triangulation with the cubulation of the spatial manifold and $SU(2)$ group elements of LQG with the corresponding $U(1)$ group elements in QRLG. Having written the geometric variables in terms of fluxes and holonomies and the phase space coordinates of the scalar field in terms of point holonomies and smeared fluxes, the quantization is straightforward and it is performed in section \ref{V}. The resulting operator is discussed in the large $j$-limit.
We point out how if proper semiclassical states are constructed for the scalar field, then the expectation value of the field contribution to the scalar constraint reproduces the classical expression. This result provides a first check on the consistency of the adopted framework. 

In this article we use the convention with metric signature $(-,+,+,+)$, gravitational coupling constant $\kappa=16\pi G$ and $c=1$. Metric tensor is defined as $g_{\mu\nu}=e^I_{\mu}e^I_{\nu}$, where $e^I_{\mu}$ are vierbein fields. Dreibeins are denoted as $e^i_a$, where lowercase latin indexes $a,b,..=1,2,3$ label coordinate on each Cauchy hypersurface constructed by ADM decomposition \cite{Arnowitt:1962hi}, while $i,j,..=1,2,3$ are su(2) internal indexes.

\section{Quantum reduced loop gravity}\label{II}

The phase space of LQG is described by holonomies of Ashtekar-Barbero connections \cite{Ashtekar:1986yd}, smeared along some curve $\gamma$,
$h_{\gamma}\!:=\mathcal{P}\exp\!\left(\int_{\gamma}A^j_a(\gamma(s))\tau^j\dot{\gamma}^a(s)\right)$
 and by fluxes of densitized triads across some surface $S$,
$E(S)\!:=\int_S n_j\epsilon_{abc}E^a_jdx^b\wedge dx^c$.
The kinematical Hilbert space of the theory is constructed as the direct sum of the space of cylindrical functions of connections along each graph $\Gamma$,
\begin{equation}
\mathcal{H}_{kin}^{(gr)}:=\bigoplus_{\Gamma}\mathcal{H}_{\Gamma}^{(gr)}\!=L_2\big(\mathcal{A},d\mu_{AL}\big),
\end{equation}
where $\mathcal{A}$ is the space of connections, $d\mu_{AL}$ denotes the Ashtekar-Lewandowski measure \cite{Ashtekar:1994mh}, while the states are cylindrical functions of all links $l_i\in\Gamma$ and they are defined as
$\Psi_{\Gamma, f}(A):=\left<A\middle|\Gamma,f\right>:=f\big( h_{l_1}(A), h_{l_2}(A), ... , h_{l_L}(A) \big)$ for some continuous function
$f:SU(2)^L\longrightarrow\mathds{C}$.

The basis states labeled with a graph $\Gamma$, with irreducible representations $D^{j_l}(h_l)$ (Wigner matrices) of spin $j$ of the holonomy along each link $l$, and with an intertwiner  $i_v$ implementing SU(2) invariance at each node $v$, are called spin network states and are given by the expression:
\begin{equation}
\Psi_{\Gamma,j_l,i_v}(h)=\left<h\middle|\{\Gamma,{j_l},{i_v}\}\right>=\prod_{v\in\Gamma}i_v\cdot\prod_lD^{j_l}(h_l),
\end{equation}
where the product $\prod_l$ extends over all the links $l$ emanating from $v$ and the $\cdot$ denotes contraction of the SU(2) indexes.

QRLG implements the restriction to diagonal spatial metric tensor and triads along some fiducial directions, along which we define some coordinates $x,y,z$. The metric tensor reads 
\begin{equation}
dl^2=a_1^2dx^2+a_2^2dy^2+a_3^2dz^2,
\end{equation}
where the three scale factors are functions of time and of all spatial coordinates. 
The graph $\Gamma$ now contains only three three kinds of links, each one being the set of links $l_i$ along a fiducial direction. Inverse densitized triads are fixed to be diagonal, 
\begin{equation}\label{diagonalmomentum}
E^i_a=p^i\delta^i_a,\quad |p^i|=\frac{a_1a_2a_3}{a_i}
\end{equation}
(indexes are not summed in this expression), and this implies a SU(2) gauge-fixing condition in the internal space. Such a gauge-fixing is realized by the projection of SU(2) group elements, which are based at links $l_i$, onto U(1) group representations obtained by stabilizing the SU(2) group along the internal directions $\vec{u}_l=\vec{u}_i$, with 
\begin{equation}
\vec{u}_1=(1,0,0)\quad \vec{u}_2=(0,1,0) \quad \vec{u}_3=(0,0,1). 
\end{equation}
The kinematical Hilbert space now reads:
\begin{equation}
^{R\!}\mathcal{H}_{kin}^{(gr)}:=\bigoplus_{\Gamma}\ \!\!^{R\!}\mathcal{H}_{\Gamma}^{(gr)}.
\end{equation}
where $\Gamma$ is a cuboidal graph, while $\!\!^{R\!}\mathcal{H}_{\Gamma}^{(gr)}$ denotes the reduced Hilbert space on a fixed (reduced) graph. The basis states in $\!\!^{R\!}\mathcal{H}_{\Gamma}^{(gr)}$ are obtained by projecting SU(2) Wigner matrices on the state of maximum or minimum magnetic number $m_l=\pm j_l$, for the angular momentum component $J_l=\vec{J}\cdot \vec{u}_l$ along the link $l$:
\begin{equation}
^{l}\!D^{j_l}_{m_lm_l}(h_l)=\left<m_l,\vec{u}_l\middle|D^{j_l}(h_l)\middle|m_l,\vec{u}_l\right>,\ \ h_l\in\text{SU(2)}.
\end{equation}
Then, reduced states, called reduced spin network states, are given by the formula:
\begin{equation}
^{R\!}\Psi_{\Gamma,m_l,i_v}(h)=\left<h\middle|\{\Gamma,m_l,i_v\}\right>
=\prod_{v\in\Gamma}\left<j_l,i_v\middle|m_l,\vec{u}_l\right>\cdot\prod_l\ \!\!^{l}\!D^{j_l}_{m_lm_l}(h_l), \ \ m_l=\pm j_l,
\end{equation}
where $\left<j_l,i_v\middle|m_l,\vec{u}_l\right>$ are reduced (one-dimensional) intertwiners.

The graphical way to construct the elements of $\!\!^{R\!}\mathcal{H}_{\Gamma}^{(gr)}$ out of those of the full theory is to replace SU(2) basis elements with the following objects
\begin{equation}
\raisebox{-1.5ex}{
\begin{tikzpicture}
\draw[l] (-1.2,0) --  node[i] {$j_l$} (-0.7,0);
\draw[l] (-0.7,-0.1) -- (-0.7,0.1);
\draw[l] (-0.5,-0.1) -- (-0.5,0.1);
\draw[l] (-0.3,0) -- (-0.5,0);
\draw[l] (0,0) circle(3mm); \node at (0.02,0) {$h_l$};
\draw[l] (0.3,0) -- (0.5,0);
\draw[l] (0.5,-0.1) -- (0.5,0.1);
\draw[l] (0.7,-0.1) -- (0.7,0.1);
\draw[l] (0.7,0) --  node[i] {$j_l$} (1.2,0);
\end{tikzpicture}
}
=
\left<j_l,m\middle|m'',\vec{u}_l\right>\left<m'',\vec{u}_l\middle|D^{j_l}(h_l)\middle|m'',\vec{u}_l\right>\left<m'',\vec{u}_l\middle|j_l,m'\right>,\quad m''=\pm j_l.
\end{equation}

Finally, reduction of canonical variables to $^{R\!}h_{l_i}$ and $^{R\!}E(S)$ is obtained by smearing along links of reduced, cuboidal graph $\Gamma$ and across surfaces $S$ perpendicular to these links, respectively.

The scalar constraint operator, neglecting the scalar curvature term, is obtained from that of LQG by considering only the euclidean part and replacing LQG operators with reduced ones. 
Its action on three-valent and six-valent nodes has been analyzed in \cite{Alesci:2013xd} and \cite{Alesci:2015nja}, respectively. 

In what follows, we will consider the generic case in which the nodes of $\Gamma$ are six-valent and we will represent the states based at each of them in the following graphical way:
\begin{equation}
\ket{\Gamma;U_{\psi}}_{\!R}
=\!\vket{
\begin{tikzpicture}
\draw[l] (-1,0) --  node[i] {$j_{x,y-1,z}^{(y)}$} (-0.2,0);
\draw[l] (-0.2,0) -- (0,0);
\draw[l] (-1,-0.1) -- (-1,0.1);
\draw[l] (-1.2,-0.1) -- (-1.2,0.1);
\draw[l] (-1.2,0) -- (-1.35,0);
\draw[l] (-1.8,0) circle(4.5mm); \node at (-1.78,0.05) {$h_{^{x\!,y\texttt{-\!}1\!,z}}^{\!(y)}$};
\draw[l] (-2.25,0) -- (-2.4,0);
\draw[l] (-2.4,-0.1) -- (-2.4,0.1);
\draw[l] (-2.6,-0.1) -- (-2.6,0.1);
\draw[l] (-3.3,0) -- node[i] {$j_{x,y-1,z}^{(y)}$} (-2.6,0);
\draw[l] (-3.6,-0.8) -- (-3.6,0.8);
\draw[l] (-3.3,0) -- (-3.8,0);
\draw[l] (0,0) --  node[i] {$j_{x,y,z}^{(y)}$} (1,0);
\draw[l] (1,-0.1) -- (1,0.1);
\draw[l] (1.2,-0.1) -- (1.2,0.1);
\draw[l] (1.2,0) -- (1.35,0);
\draw[l] (1.8,0) circle(4.5mm); \node at (1.82,0.05) {$h_{^{x\!,y\!,z}}^{\!(y)}$};
\draw[l] (2.25,0) -- (2.4,0);
\draw[l] (2.4,-0.1) -- (2.4,0.1);
\draw[l] (2.6,-0.1) -- (2.6,0.1);
\draw[l] (2.6,0) -- node[i] {$j_{x,y,z}^{(y)}$} (3.6,0);
\draw[l] (3.6,-0.8) -- (3.6,0.8);
\draw[l] (3.6,0) -- (3.8,0);
\draw[l] (0,0.7) --  node[j] {$j_{x,y,z}^{(z)}$} (0,1);
\draw[l] (0,0) -- (0,0.7);
\draw[l] (-0.1,1) -- (0.1,1);
\draw[l] (-0.1,1.2) -- (0.1,1.2);
\draw[l] (0,1.2) -- (0,1.35);
\draw[l] (0,1.8) circle(4.5mm); \node at (0.02,1.85) {$h_{^{x\!,y\!,z}}^{\!(z)}$};
\draw[l] (0,2.25) -- (0,2.4);
\draw[l] (-0.1,2.4) -- (0.1,2.4);
\draw[l] (-0.1,2.6) -- (0.1,2.6);
\draw[l] (0,2.6) --  node[i] {$j_{x,y,z}^{(z)}$} (0,3.6);
\draw[l] (0,3.6) -- (0,3.8);
\draw[l] (-0.8,3.6) -- (0.8,3.6);
\draw[l] (0,-0.4) -- (0,0);
\draw[l] (0,-1) --  node[i] {$j_{x,y,z-1}^{(z)}$} (0,-0.4);
\draw[l] (-0.1,-1) -- (0.1,-1);
\draw[l] (-0.1,-1.2) -- (0.1,-1.2);
\draw[l] (0,-1.2) -- (0,-1.35);
\draw[l] (0,-1.8) circle(4.5mm); \node at (0.02,-1.75) {$h_{^{x\!,y\!,z\texttt{-\!}1}}^{\!(z)}$};
\draw[l] (0,-2.25) -- (0,-2.4);
\draw[l] (-0.1,-2.4) -- (0.1,-2.4);
\draw[l] (-0.1,-2.6) -- (0.1,-2.6);
\draw[l] (0,-3.6) --  node[i] {$j_{x,y,z-1}^{(z)}$} (0,-2.6);
\draw[l] (0,-3.8) -- (0,-3.6);
\draw[l] (0.8,-3.6) --  (-0.8,-3.6);
\node at (-1.8,1.6) {$\atopp{x\ direction}{\odot}$};
\node at (-1.8,1) {$\atopp{-x\ direction\ \ }{\otimes}$};
\end{tikzpicture}
},
\end{equation}
where the node $v_{x,y,z}$ is placed at $(x,y,z)$, 
while the symbol $j_{i+n,j,k}^{(i)}$ denotes the spin number attached to a link along the $i$-axis, beginning at the node $\{i\!+\!n,j,k\}$ and ending at the node $\{i\!+\!n\!+\!1,j,k\}$.  We assumed the right-handed orientation of links, {\it i.e.} the link with the spin number $j_{i+n,j,k}^{(i)}$ is outgoing from the node $i+n,j,k$, while $j_{i+n-1,j,k}^{(i)}$ is ingoing to the same node. 

In what follows, we will need the expression of the powers of the volume operator $\hat{\mathbf{V}}$, which acts diagonally \cite{Alesci:2013xd} as follows:
\begin{equation}\label{volume}
\left(\hat{\mathbf{V}}(v_{x,y,z}) \right)^{\!\!n}\ket{\Gamma;U_{\psi}}_{\!R}
=\left(8\pi\gamma l_P^2\right)^{\!\frac{3}{2}n}
\left(
\frac{j_{x-1,y,z}^{(x)}\!+j_{x,y,z}^{(x)}}{2}\times\frac{j_{x,y-1,z}^{(y)}\!+j_{x,y,z}^{(y)}}{2}\times\frac{j_{x,y,z-1}^{(z)}\!+j_{x,y,z}^{(z)}}{2}
\right)^{\!\frac{n}{2}}
\ket{\Gamma;U_{\psi}}_{\!R},
\end{equation}
where $\gamma$ denotes the Immirzi parameter.


\section{Loop framework for quantum scalar field}\label{III}

The action of the scalar field minimally coupled to gravity reads:
\begin{equation}\label{action}
S^{(\phi)}=\frac{1}{2\lambda}\int_M\!\!\!\!d^{4}x\sqrt{-g}\big(g^{\mu\nu}(\partial_{\mu}\phi)(\partial_{\nu}\phi)-V(\phi)\big),
\end{equation}
where $\lambda$ is the coupling constant of dimension {\bf$1/\hbar$} and $g$ is the determinant of four-dimensional metric tensor.

The Legendre transform gives the following Hamiltonian:
\begin{equation}\label{Hamiltonian}
H^{(\phi)}\!=\!\int_{\Sigma_t}\!\!\!\!d^3x\Bigg(N^a\pi\partial_a\phi
+N\bigg(\frac{\lambda}{2\sqrt{q}}\pi^2+\frac{\sqrt{q}}{2\lambda}q^{ab}\partial_a\phi\partial_b\phi+\frac{\sqrt{q}}{2\lambda}V(\phi)\bigg)\Bigg)=\!\int_{\Sigma_t}\!\!\!\!d^3x\big(N^a\mathcal{V}_a^{(\phi)}+NH^{(\phi)}_{\text{sc}}\big),
\end{equation}
$N$ and $N^a$ being the lapse function and the shift vector, respectively, while $\mathcal{V}_a^{(\phi)}$ and $H^{(\phi)}_{\text{sc}}$ are the contributions of the scalar field to the vector and scalar constraints. $q$ denotes the determinant of spatial metric and $\pi$ is the conjugate momentum to the scalar field.

The total vector constraint is the sum of $\mathcal{V}_a^{(\phi)}:=\pi\partial_a\phi$ plus the gravitational part and it generates diffeomorphisms.

The second term $H^{(\phi)}_{\text{sc}}[N]$ in the expression (\ref{Hamiltonian}), which is the field contribution to the smeared scalar constraint encodes all information about the dynamics of the scalar field in the diffeomorphisms invariant phase-space. It can be written as follows 
\begin{equation}\label{Hamiltonianconstraint}
H^{(\phi)}_{\text{sc}}[N]:=\!\int_{\Sigma_t}\!\!\!\!d^3xN\bigg(\frac{\lambda}{2\sqrt{q}}\pi^2+\frac{\sqrt{q}}{2\lambda}q^{ab}\partial_a\phi\partial_b\phi+\frac{\sqrt{q}}{2\lambda}V(\phi)\bigg):=H^{(\phi)}_{kin}[N]+H^{(\phi)}_{der}[N]+H^{(\phi)}_{pot}[N],
\end{equation}
where we split it into three parts, the kinetic, derivative and potential ones, {\it i.e.}
\begin{align}
&H^{(\phi)}_{kin}[N]=\!\int_{\Sigma_t}\!\!\!\!d^3xN\bigg(\frac{\lambda}{2\sqrt{q}}\pi^2\bigg)\label{kinetic}\\
&H^{(\phi)}_{der}[N]=\!\int_{\Sigma_t}\!\!\!\!d^3xN\bigg(\frac{\sqrt{q}}{2\lambda}q^{ab}\partial_a\phi\partial_b\phi\bigg)\label{derivative}\\
&H^{(\phi)}_{pot}[N]=\!\int_{\Sigma_t}\!\!\!\!d^3xN\bigg(\frac{\sqrt{q}}{2\lambda}V(\phi)\bigg).\label{potential}
\end{align}

We quantize the system of gravity and the scalar field by adapting the procedure described in \cite{Thiemann:1997rt,Thiemann:1997rq} for LQG to the case of QRLG. Hence, the total Hilbert space is the direct product of that for gravity times that for $\phi$ 
\begin{equation}
\mathcal{H}_{kin}^{(tot)}={}^{R\!}\mathcal{H}_{kin}^{(gr)}\!\otimes\mathcal{H}_{kin}^{(\phi)},
\end{equation}
the latter being the following Hilbert space
\begin{equation}
\mathcal{H}_{kin}^{(\phi)}:=\overline{\big\{a_1U_{\psi_1}+...+a_nU_{\psi_n}\!:\ a_i\in\mathds{C},\, n\in\mathds{N},\, \psi_i\in\mathds{R}\big\}},
\end{equation}
where the states are defined as
\begin{equation}
U_\psi=e^{i\sum_{v\in\Sigma}\psi_v \phi_v}=\ket{U_\psi}  
\end{equation}
with the normalization provided by the scalar product
\begin{equation}
\bracket{U_{\psi}}{U_{\psi'}}:=\delta_{\psi,\psi'}.
\end{equation}
The basic variables act as follows:
\begin{equation}\label{canonicaloperators-1}
\hat{U}_\psi\ket{U_{\psi'}}=\ket{U_{\psi+\psi'}}, \ \ 
\hat{\Pi}(V)\ket{U_{\psi}}=\hbar\sum_{v\in V}\psi_v\ket{U_{\psi}}\,,
\end{equation}
$\Pi(V)$ being the scalar field momentum smeared over the volume $V\subseteq\Sigma$. 
One can also define single-point states 
\begin{equation}
\ket{v;U_{\psi}}:=e^{i\psi_v\phi_v}\,,
\end{equation}
for which the scalar product reads 
\begin{equation}
\bracket{w;U_{\psi}}{v;U_{\psi'}}:=\delta_{w,v}\delta_{\psi,\psi'}\,,
\end{equation}
while basic operators act as follows:
\begin{equation}\label{canonicaloperators}
e^{i\psi_w\hat{\phi}_w}\ket{v;U_{\psi}}=e^{i\psi_w\phi_w}\ket{v;U_{\psi}}=\ket{v\cup w;U_{\psi}}, \ \ 
\hat{\Pi}(v)\ket{v;U_{\psi}}=-i\hbar\frac{\partial}{\partial\phi(v)}\ket{v;U_{\psi}}=\hbar\psi_v\ket{v;U_{\psi}}\,,
\end{equation}
and one can define the smeared field around a point $v$ via
\begin{equation}\label{smearedmomentum}
\Pi(v):=\int\!\!d^3u \chi_{\varepsilon}(v,u)\pi(u).
\end{equation}
where one introduces the characteristic function $\chi_{\varepsilon}(v,u)$ of the box $B_{\varepsilon}(v)$ centered in $v$ with coordinate volume $\varepsilon^3$, precisely
\begin{equation}
\mathbf{V}\big(B_{\varepsilon}(v)\big):=\mathbf{V}(v,\varepsilon)=\varepsilon^3\sqrt{q}(v)+O(\varepsilon^4),
\end{equation}
which allows to smear a function at the point $v$, around infinitesimal neighborhood, such that
\begin{equation}\label{characteristic}
f(v)=\int\!\!d^3u\,\delta^3(v-u)f(u)=\lim_{\varepsilon\to0}\frac{1}{\varepsilon^3}\!\int\!\!d^3u\,\chi_{\varepsilon}(v,u)f(u).
\end{equation}

This way, the commutators are finite
\begin{subequations}
\begin{align}
\big\{\phi(x),\phi(y)\big\}=\big\{\Pi(x),\Pi(y)\big\}&=0,\\
\big\{\phi(x),\Pi(y)\big\}&=\chi_{\varepsilon}(x,y).
\end{align}
\end{subequations}

The full Hilbert space $\mathcal{H}_{kin}^{(\phi)}:=L_2\left(\bar{\mathds{R}}_{\text{Bohr}}{}^\Sigma\right)$ can be obtained from the  single-point one $L_2\big(\bar{\mathds{R}}_{\text{Bohr}}\big)$, where $\bar{\mathds{R}}_{\text{Bohr}}$ denotes the Bohr compactification of a line and the Bohr measure is defined as
\begin{equation}
\int_{\bar{\mathds{R}}_{\text{Bohr}}}d\mu_{\text{Bohr}}(\phi)e^{i\psi_v\phi_v}=\delta_{0,v}\,.
\end{equation}

This method for treating a scalar field on a lattice realizes a polymer representation in the momentum polarization (also called point-holonomy representation) \cite{Ashtekar:2002sn,Ashtekar:2002vh,Kaminski:2005nc,Kaminski:2006ta}.
In QRLG, the only difference with respect to the scalar field quantization in full LQG is just the restriction to cuboidal lattices. 

The states in the total Hilbert space $\mathcal{H}_{kin}^{(tot)}$ are described in the following way:
\begin{equation}\label{state}
\ket{\Gamma;m_l,i_v;U_{\psi}}_{\!R}=\!
\ket{\Gamma;m_l,i_v}_{\!R}\otimes\ket{\Gamma;U_{\psi}}_{\!R}
=\!\vket{
\begin{tikzpicture}
\draw[l] (-1,0) --  node[i] {$j_{x,y-1,z}^{(y)}$} (-0.2,0);
\draw[l] (-0.2,0) -- (0,0);
\draw[l] (-1,-0.1) -- (-1,0.1);
\draw[l] (-1.2,-0.1) -- (-1.2,0.1);
\draw[l] (-1.2,0) -- (-1.35,0);
\draw[l] (-1.8,0) circle(4.5mm); \node at (-1.78,0.05) {$h_{^{x\!,y\texttt{-\!}1\!,z}}^{\!(y)}$};
\draw[l] (-2.25,0) -- (-2.4,0);
\draw[l] (-2.4,-0.1) -- (-2.4,0.1);
\draw[l] (-2.6,-0.1) -- (-2.6,0.1);
\draw[l] (-3.3,0) -- node[i] {$j_{x,y-1,z}^{(y)}$} (-2.6,0);
\draw[l] (-3.6,-0.8) -- (-3.6,0.8);
\draw[l] (-3.3,0) -- (-3.8,0);
\node at (-2.55,-0.55) {$e^{i\psi_{x\!,y\texttt{-\!}1\!,z}\phi_{x\!,y\texttt{-\!}1\!,z}}$};
\draw[l] (0,0) --  node[i] {$j_{x,y,z}^{(y)}$} (1,0);
\draw[l] (1,-0.1) -- (1,0.1);
\draw[l] (1.2,-0.1) -- (1.2,0.1);
\draw[l] (1.2,0) -- (1.35,0);
\draw[l] (1.8,0) circle(4.5mm); \node at (1.82,0.05) {$h_{^{x\!,y\!,z}}^{\!(y)}$};
\draw[l] (2.25,0) -- (2.4,0);
\draw[l] (2.4,-0.1) -- (2.4,0.1);
\draw[l] (2.6,-0.1) -- (2.6,0.1);
\draw[l] (2.6,0) -- node[i] {$j_{x,y,z}^{(y)}$} (3.6,0);
\draw[l] (3.6,-0.8) -- (3.6,0.8);
\draw[l] (3.6,0) -- (3.8,0);
\node at (3.48,-0.5) {$e^{i\psi_{x\!,y\texttt{+\!}1\!,z}\phi_{x\!,y\texttt{+\!}1\!,z}}$};
\node at (0.87,-0.5) {$e^{i\psi_{x\!,y\!,z}\phi_{x\!,y\!,z}}$};
\draw[l] (0,0.7) --  node[j] {$j_{x,y,z}^{(z)}$} (0,1);
\draw[l] (0,0) -- (0,0.7);
\draw[l] (-0.1,1) -- (0.1,1);
\draw[l] (-0.1,1.2) -- (0.1,1.2);
\draw[l] (0,1.2) -- (0,1.35);
\draw[l] (0,1.8) circle(4.5mm); \node at (0.02,1.85) {$h_{^{x\!,y\!,z}}^{\!(z)}$};
\draw[l] (0,2.25) -- (0,2.4);
\draw[l] (-0.1,2.4) -- (0.1,2.4);
\draw[l] (-0.1,2.6) -- (0.1,2.6);
\draw[l] (0,2.6) --  node[i] {$j_{x,y,z}^{(z)}$} (0,3.6);
\draw[l] (0,3.6) -- (0,3.8);
\draw[l] (-0.8,3.6) -- (0.8,3.6);
\node at (1.05,3.35) {$e^{i\psi_{x\!,y\!,z\texttt{+\!}1}\phi_{x\!,y\!,z\texttt{+\!}1}}$};%
\draw[l] (0,-0.4) -- (0,0);
\draw[l] (0,-1) --  node[i] {$j_{x,y,z-1}^{(z)}$} (0,-0.4);
\draw[l] (-0.1,-1) -- (0.1,-1);
\draw[l] (-0.1,-1.2) -- (0.1,-1.2);
\draw[l] (0,-1.2) -- (0,-1.35);
\draw[l] (0,-1.8) circle(4.5mm); \node at (0.02,-1.75) {$h_{^{x\!,y\!,z\texttt{-\!}1}}^{\!(z)}$};
\draw[l] (0,-2.25) -- (0,-2.4);
\draw[l] (-0.1,-2.4) -- (0.1,-2.4);
\draw[l] (-0.1,-2.6) -- (0.1,-2.6);
\draw[l] (0,-3.6) --  node[i] {$j_{x,y,z-1}^{(z)}$} (0,-2.6);
\draw[l] (0,-3.8) -- (0,-3.6);
\draw[l] (0.8,-3.6) --  (-0.8,-3.6);
\node at (1.05,-3.35) {$e^{i\psi_{x\!,y\!,z\texttt{-\!}1}\phi_{x\!,y\!,z\texttt{-\!}1}}$};
\node at (-1.8,1.6) {$\atopp{x\ direction}{\odot}$};
\node at (-1.8,1) {$\atopp{-x\ direction\ \ }{\otimes}$};
\end{tikzpicture}\!\!
}.\!
\end{equation}
The scalar field state is given by attaching at the each node $v_{p,q,r}\!\in\!\Gamma$ the point holonomy $e^{i\psi_{p,q,r}\phi_{p,q,r}}$ with the real coefficient $\psi_{p,q,r}$, while the gravity part is described by the spin numbers $j_{i,j,k}^{(i)}$ at the associated links $l_{i,j,k}^{(i)}$ and the reduced interwiners at nodes.

\section{Regularization of scalar constraint}\label{IV}

The quantization of the scalar part of the scalar constraint requires a regularization of the first two terms in the expression (\ref{Hamiltonianconstraint}). This is done going back to the classical phase space and rewriting \eqref{Hamiltonianconstraint} in terms of holonomies and fluxes for the gravitational part, point holonomies and smeared momenta for the terms involving the scalar field.

The gravitational part is regularized by the method developed in LQG \cite{Thiemann:1996aw}, restricted to a cuboidal graph \cite{Alesci:2013xd,Alesci:2015nja}. Basically, the idea is just to replace the triangulation of the spatial metric with a cubulation. In this description, matter coupled to a dynamical spacetime is regularized by a reduction to matter fields coupled to a dynamical lattice (nodes and links). Alternatively, in a dual picture, matter fields are coupled to dynamics of granulated space (volumes and areas of chunks of space) \cite{covariant}.

In order to regularize $H^{(\phi)}_{kin}$ and $H^{(\phi)}_{der}$, we mimic the procedure given in \cite{Thiemann:1997rt,Thiemann:1997rq} for full LQG in the presence of a scalar field. At first, smearing momenta and inserting the expression $\frac{e^2}{q}=1$, where $e$ is a determinant of triad $e^i_a$, one can write the following expression for the kinetic term
\eqref{kinetic}
\begin{equation}\label{classicalkin1}
\begin{split}
H^{(\phi)}_{kin}&=\frac{\lambda}{2}\lim_{\varepsilon\to0}
\int\!\!d^3x\,N(x)\pi(x)\int\!\!d^3y\pi(y)\int\!\!d^3t\frac{e}{\big(\mathbf{V}(t,\varepsilon)\big)^{\!\frac{3}{2}}}
\int\!\!d^3u\frac{e}{\big(\mathbf{V}(u,\varepsilon)\big)^{\!\frac{3}{2}}}
\chi_{\varepsilon}(x,y)\chi_{\varepsilon}(t,x)\chi_{\varepsilon}(u,y),
\end{split}
\end{equation}
where it has been used the definition of the volume $\mathbf{V}(R)$ of the region $R$
\begin{equation}
\mathbf{V}(R)=\int_{\!R}\!\!d^3x\sqrt{q}.
\end{equation}

Then, to remove the denominators in formula (\ref{classicalkin1}), we use Thiemann's trick \cite{Thiemann:1996aw}. Using the Poisson brackets between connection and volume, one can derive the following relation
\begin{equation}\label{trick1}
e^i_a(x)=2\frac{\delta\mathbf{V}(R)}{\delta E_i^a}=\frac{2}{n\big(\mathbf{V}(R)\big)^{\!n-1}}\frac{\delta\big(\mathbf{V}(R)\big)^{\!n}}{\delta E_i^a}
=\frac{4}{n\gamma\kappa\big(\mathbf{V}(R)\big)^{\!n-1}}\Big\{A^i_a(x),\big(\mathbf{V}(R)\big)^{\!n}\Big\}
\end{equation}
and use it to write 
\begin{equation}\label{smearing1}
\begin{split}
\int\!\!d^3t\frac{e}{\big(\mathbf{V}(t,\varepsilon)\big)^{\!\frac{3}{2}}}
=&\,\frac{1}{3!}\epsilon_{ijk}\!\int\!\!\frac{e^i}{\big(\mathbf{V}(t,\varepsilon)\big)^{\!\frac{1}{2}}}
\wedge\frac{e^j}{\big(\mathbf{V}(t,\varepsilon)\big)^{\!\frac{1}{2}}}\wedge
\frac{e^k}{\big(\mathbf{V}(t,\varepsilon)\big)^{\!\frac{1}{2}}}
=\\
=&\,\frac{1}{6}\!\left(\frac{8}{\gamma\kappa}\right)^{\!\!3}\!\epsilon_{ijk}\!\int\!
\left\{A^i(t),\big(\mathbf{V}(t,\varepsilon)\big)^{\!\frac{1}{2}}\right\}
\!\wedge\!\left\{A^j(t),\big(\mathbf{V}(t,\varepsilon)\big)^{\!\frac{1}{2}}\right\}\!\wedge\!
\left\{A^k(t),\big(\mathbf{V}(t,\varepsilon)\big)^{\!\frac{1}{2}}\right\},
\end{split}
\end{equation}
which can be inserted into (\ref{classicalkin1}), thus giving
\begin{equation}\label{classicalkin2}
\begin{split}
H^{(\phi)}_{kin}
=&\,\frac{\lambda}{2}\frac{2^{16}}{3^2(\gamma\kappa)^6}\epsilon_{ijk}\epsilon_{lmn}\lim_{\varepsilon\to0}
\int\!\!d^3r\,N(r)\pi(r)\,\chi_{\varepsilon}(r,x)\int\!\!d^3s\,\pi(s)\chi_{\varepsilon}(y,s)\times\\
&\times
\!\int\!
\left\{A^i(t),\big(\mathbf{V}(t,\varepsilon)\big)^{\!\frac{1}{2}}\right\}
\!\wedge\!\left\{A^j(t),\big(\mathbf{V}(t,\varepsilon)\big)^{\!\frac{1}{2}}\right\}\!\wedge\!
\left\{A^k(t),\big(\mathbf{V}(t,\varepsilon)\big)^{\!\frac{1}{2}}\right\}\!
\times\\
&\times
\!\int\!
\left\{A^l(u),\big(\mathbf{V}(u,\varepsilon)\big)^{\!\frac{1}{2}}\right\}
\!\wedge\!\left\{A^m(u),\big(\mathbf{V}(u,\varepsilon)\big)^{\!\frac{1}{2}}\right\}\!\wedge\!
\left\{A^n(u),\big(\mathbf{V}(u,\varepsilon)\big)^{\!\frac{1}{2}}\right\}\!
\times\\
&\times
\chi_{\varepsilon}(r,s)\chi_{\varepsilon}(t,r)\chi_{\varepsilon}(u,s).
\end{split}
\end{equation}

A method of discretization of the scalar constraint via a triangularization of the spatial manifold has been developed in \cite{Thiemann:1996aw} for pure gravity and then applied in the presence of a scalar field \cite{Thiemann:1997rt}. The idea is to replace the integration over the spatial hypersurface $\int_{\Sigma}$ with the sum over over all ordered tetrahedra. 
Hence, the sum over tetrahedra becomes the sum over all the nodes $v$ of the triangulation and over all the tetrahedra $\Delta_{l,l',l''}$ created by triples of links $\{l,l',l''\}$ emanating form $v$. 
Given a cubulation, each node $v$ is always surrounded by three pairs of links, oriented along fixed perpendicular directions. They always create eight tetrahedra around the node and it is worth nothing that for each tetrahedron, the remaining seven ones coincide with the seven "virtual" tetrahedra, which must be constructed to triangulate any cuboidal or non-cuboidal lattice.

Finally, integration over each tetrahedron, $\int_{\Delta_{l,l'\!,l''}}$, turns into the sum over the eight possibilities for choosing a triple of perpendicular links $\{l,l',l''\}$ among each tetrahedron of the triangulation $\Delta(v)$ around the node $v$.

Using the triangulation procedure one gets the result 
\begin{equation}\label{regularization1}
\begin{split}
&\,\epsilon_{ijk}\epsilon^{abc}\!\int\!\!d^3t
\left\{A_a^i(t),\big(\mathbf{V}(t,\varepsilon)\big)^{\!\frac{1}{2}}\right\}
\!\left\{A_b^j(t),\big(\mathbf{V}(t,\varepsilon)\big)^{\!\frac{1}{2}}\right\}\!
\left\{A_c^k(t),\big(\mathbf{V}(t,\varepsilon)\big)^{\!\frac{1}{2}}\right\}
\!\chi_{\varepsilon}(t,w)
=\\
=&\,\epsilon_{ijk}\,\varepsilon^3\!\!\!\!\sum_{v\in\mathbf{V}(\Gamma)}\sum_{v_{l,l'\!,l''}}\!\int_{\!\Delta_{l,l'\!,l''}}\!\!\!\!\!\!\!
\left\{A_a^i(v),\big(\mathbf{V}(v_{l,l'\!,l''},\varepsilon)\big)^{\!\frac{1}{2}}\right\}\!\delta^a_{l}
\!\wedge\!\left\{A_b^j(v),\big(\mathbf{V}(v_{l,l'\!,l''},\varepsilon)\big)^{\!\frac{1}{2}}\right\}\!\delta^b_{l'}\wedge\\
&\hspace{30 mm}\wedge\!\left\{A_c^k(v),\big(\mathbf{V}(v_{l,l'\!,l''},\varepsilon)\big)^{\!\frac{1}{2}}\right\}\!\delta^c_{l''}\,\delta_{v,w}
\approx\\
\approx&\,2^3\!\!\!\!\sum_{v\in\mathbf{V}(\Gamma)}\sum_{\Delta(v)}\epsilon_{ijk}\epsilon_{pqr}\,\text{tr}
\bigg(\!
\tau^ih_{l_p(\Delta)}^{-1}\left\{\Big(\mathbf{V}\big(\Delta(v),\varepsilon\big)\Big)^{\!\frac{1}{2}},\,h_{l_p(\Delta)}\right\}
\!\bigg)\,
\text{tr}
\bigg(\!
\tau^jh_{l_q(\Delta)}^{-1}\left\{\Big(\mathbf{V}\big(\Delta(v),\varepsilon\big)\Big)^{\!\frac{1}{2}},\,h_{l_q(\Delta)}\right\}
\!\bigg)\times\\
&\hspace{27 mm}\times\text{tr}
\bigg(\!
\tau^kh_{l_r(\Delta)}^{-1}\left\{\Big(\mathbf{V}\big(\Delta(v),\varepsilon\big)\Big)^{\!\frac{1}{2}},\,h_{l_r(\Delta)}\right\}
\!\bigg)\delta_{v,w},
\end{split}
\end{equation}
where $h_{l(\Delta)}$ is the SU(2) holonomy operator in the fundamental representation \cite{Thiemann:1996aw}, tr denotes the trace over SU(2) algebra and $\tau^j=-\frac{i}{2}\sigma^j$, while $\sigma^j$ are Pauli matrices. The summations $\sum_{v\in\mathbf{V}(\Gamma)}$ and $\sum_{\Delta(v)}$ extend over all nodes of the cubulation and over all the tetrahedra around each node, respectively. In the last line the following expansion
\begin{equation}\label{trick2}
\,\text{tr}\big(\tau^ih_{l_a}^{-1}\big\{\mathbf{V}^{n}(R),h_{l_a}\big\}\big)
=-\,\text{tr}\big(\tau^i\varepsilon\big\{A_a,\mathbf{V}^{n}(R)\big\}+O(\varepsilon^2)\big)
\approx\frac{1}{2}\varepsilon\big\{A_a^i,\mathbf{V}^{n}(R)\big\}
\end{equation}
 has been applied. As a result, the kinetic part of scalar Hamiltonian constraint operator reads
\begin{equation}\label{kin2}
\begin{split}
H^{(\phi)}_{kin}=
&\frac{2^{21}\lambda}{3^2(\gamma\kappa)^6}\lim_{\varepsilon\to0}
\!\!\!\!\sum_{v,v'\in\mathbf{V}(\Gamma)}\!\!\!\!\!\!N(v)\Pi(v)\Pi(v')\,\chi_{\varepsilon}(v,v')
\sum_{\Delta(v)}\,\sum_{\Delta'(v')}\!\!\epsilon_{ijk}\epsilon_{pqr}\epsilon_{lmn}\epsilon_{stu}\times\\
&\times\text{tr}\bigg(\!
\tau^ih_{l_p(\Delta)}^{-1}\left\{\Big(\mathbf{V}\big(\Delta(v),\varepsilon\big)\Big)^{\!\frac{1}{2}},\,h_{l_p(\Delta)}\right\}
\!\bigg)\,
\text{tr}\bigg(\!
\tau^jh_{l_q(\Delta)}^{-1}\left\{\Big(\mathbf{V}\big(\Delta(v),\varepsilon\big)\Big)^{\!\frac{1}{2}},\,h_{l_q(\Delta)}\right\}
\!\bigg)\times\\
&\times\text{tr}\bigg(\!
\tau^kh_{l_r(\Delta)}^{-1}\left\{\Big(\mathbf{V}\big(\Delta(v),\varepsilon\big)\Big)^{\!\frac{1}{2}},\,h_{l_r(\Delta)}\right\}
\!\bigg)\,\text{tr}\bigg(\!
\tau^lh_{l_s(\Delta\!')}^{-1}\left\{\Big(\mathbf{V}\big(v'\!(\Delta\!'),\varepsilon\big)\Big)^{\!\frac{1}{2}},\,h_{l_s(\Delta\!')}\right\}
\!\bigg)\times\\
&\times\text{tr}\bigg(\!
\tau^mh_{l_t(\Delta\!')}^{-1}\left\{\Big(\mathbf{V}\big(v'\!(\Delta\!'),\varepsilon\big)\Big)^{\!\frac{1}{2}},\,h_{l_t(\Delta\!')}\right\}
\!\bigg)\,
\text{tr}\bigg(\!
\tau^nh_{l_u(\Delta\!')}^{-1}\left\{\Big(\mathbf{V}\big(v'\!(\Delta\!'),\varepsilon\big)\Big)^{\!\frac{1}{2}},\,h_{l_u(\Delta\!')}\right\}
\!\bigg)\,,\!\!
\end{split}
\end{equation}
where \eqref{smearedmomentum} has been used. 

The derivative term (\ref{derivative}) is regularized by the same method, applying the identity 
\begin{equation}
\sqrt{q}q^{ab}=\frac{1}{4\sqrt{q}}\epsilon_{ijk}\epsilon^{acd}e^j_ce^k_d\epsilon^i_{\,lm}\epsilon^{bef}e^l_ee^m_f,
\end{equation}
and it can be written
\begin{equation}\label{classicalder1}
\begin{split}
H^{(\phi)}_{der}=&\,\frac{1}{8\lambda}\lim_{\varepsilon\to0}
\int\!\!d^3x\,N(x)\epsilon_{ijk}\epsilon^{acd}
\partial_a\phi(x)\frac{e^j_c}{\big(\mathbf{V}(x,\varepsilon)\big)^{\frac{1}{4}}}\frac{e^k_d}{\big(\mathbf{V}(x,\varepsilon)\big)^{\frac{1}{4}}}\times\\
&\times\!\int\!\!d^3y\,\epsilon^i_{\,lm}\epsilon^{bef}
\partial_b\phi(y)\frac{e^l_e}{\big(\mathbf{V}(y,\varepsilon)\big)^{\frac{1}{4}}}\frac{e^m_f}{\big(\mathbf{V}(y,\varepsilon)\big)^{\frac{1}{4}}}
\chi_{\varepsilon}(x,y),
\end{split}
\end{equation}
where the determinant of metric has been smeared and changed into a volume.

Next, using again the Thiemann's trick (\ref{trick1}) one gets the formula:
\begin{equation}\label{smearing2}
\begin{split}
\epsilon_{ijk}\epsilon^{acd}\!\!\!\int\!\!d^3x\,\partial_a\phi(x)
\frac{e^j_ce^k_d}{\big(\mathbf{V}(y,\varepsilon)\big)^{\!\frac{1}{2}}}
=&\,
\epsilon_{ijk}\!\!\int\!\!\partial\phi(x)
\wedge\frac{e^j}{\big(\mathbf{V}(x,\varepsilon)\big)^{\!\frac{1}{4}}}\wedge
\frac{e^k}{\big(\mathbf{V}(x,\varepsilon)\big)^{\!\frac{1}{4}}}
=\\
=&
\left(\!\frac{16}{3\gamma\kappa}\!\right)^{\!\!2}\!\epsilon_{ijk}\!\int\!\!
\partial\phi(x)\!
\wedge\!\left\{A^j(x),\big(\mathbf{V}(x,\varepsilon)\big)^{\!\frac{3}{4}}\right\}\!\wedge\!
\left\{A^k(x),\big(\mathbf{V}(x,\varepsilon)\big)^{\!\frac{3}{4}}\right\},
\end{split}
\end{equation}
which applied to the expression (\ref{classicalder1}) leads to the following result:
\begin{equation}\label{classicalder2}
\begin{split}
H^{(\phi)}_{der}
=&\,\frac{1}{8\lambda}\frac{2^{16}}{3^4(\gamma\kappa)^4}\epsilon_{ijk}\epsilon^i_{\,lm}\lim_{\varepsilon\to0}
\int\!\!N(x)\,
\partial\phi(x)\!
\wedge\!\left\{A^j(x),\big(\mathbf{V}(x,\varepsilon)\big)^{\!\frac{3}{4}}\right\}\!\wedge\!
\left\{A^k(x),\big(\mathbf{V}(x,\varepsilon)\big)^{\!\frac{3}{4}}\right\}\!
\times\\
&\times\!\int\!\!
\partial\phi(y)\!
\wedge\!\left\{A^j(y),\big(\mathbf{V}(y,\varepsilon)\big)^{\!\frac{3}{4}}\right\}\!\wedge\!
\left\{A^k(y),\big(\mathbf{V}(y,\varepsilon)\big)^{\!\frac{3}{4}}\right\}
\!\chi_{\varepsilon}(x,y).
\end{split}
\end{equation}

By using the same discretization adopted for the kinetic term (\ref{regularization1}), one gets
\begin{equation}\label{regularization2}
\begin{split}
&\,\epsilon_{ijk}\epsilon^{acd}\!\int\!\!d^3x\,
\partial_a\phi(x)
\!\left\{A^j(x),\big(\mathbf{V}(x,\varepsilon)\big)^{\!\frac{3}{4}}\right\}\!
\left\{A^k(x),\big(\mathbf{V}(x,\varepsilon)\big)^{\!\frac{3}{4}}\right\}
\!\chi_{\varepsilon}(x,y)=\\
=&\,\epsilon_{ijk}\,\varepsilon^3\!\!\!\!\sum_{v\in\mathbf{V}(\Gamma)}\sum_{v_{l,l',l''}}\!\int_{\Delta_{l,l',l''}}
\!\!\!
\partial_a\phi(v)\,\delta_{l}^a
\!\wedge\!\left\{A_c(v),\big(\mathbf{V}(v_{l,l',l''},\varepsilon)\big)^{\!\frac{3}{4}}\right\}\!
\delta_{l'}^c\wedge\!\left\{A_d(v),\big(\mathbf{V}(v_{l,l',l''},\varepsilon)\big)^{\!\frac{3}{4}}\right\}\!\delta_{l''}^d\,
\delta_{x,v}=\\
\!\approx&\,2^2\,\varepsilon\!\!\!\!\sum_{v\in\mathbf{V}(\Gamma)}\sum_{\Delta(v)}\!\epsilon_{ijk}\epsilon_{pqr}\,
\partial_p\phi(v)\,\text{tr}\bigg(\!
\tau^jh_{l_q(\Delta)}^{-1}\left\{\Big(\mathbf{V}\big(\Delta(v),\varepsilon\big)\Big)^{\!\frac{3}{4}},\,h_{l_q(\Delta)}\right\}
\!\bigg)\,\text{tr}\bigg(\!
\tau^kh_{l_r(\Delta)}^{-1}\left\{\Big(\mathbf{V}\big(\Delta(v),\varepsilon\big)\Big)^{\!\frac{3}{4}},\,h_{l_r(\Delta)}\right\}
\!\bigg)\delta_{x,v},\!\!\!\!\!\!
\end{split}
\end{equation}
and for $H^{(\phi)}_{der}$
\begin{equation}
\begin{split}\label{der}
\!\!\!H^{(\phi)}_{der}
=&\,\frac{2^{17}}{3^4\lambda\,(\gamma\kappa)^4}\lim_{\varepsilon\to0}\varepsilon^2
\!\!\!\!\!\!\sum_{v,v'\in\mathbf{V}(\Gamma)}\!\!\!\!\!\!N(v)\sum_{\Delta(v)}\,\sum_{\Delta'(v')}\!\!\epsilon_{ijk}\epsilon_{pqr}\epsilon^i_{\,lm}\epsilon_{stu}\,\delta_{v,v'}
\times\\
&\times
\partial_p\phi(v)\,\text{tr}\bigg(\!
\tau^jh_{l_q(\Delta)}^{-1}\left\{\Big(\mathbf{V}\big(\Delta(v),\varepsilon\big)\Big)^{\!\frac{3}{4}},\,h_{l_q(\Delta)}\right\}
\!\bigg)\,\text{tr}\bigg(\!
\tau^kh_{l_r(\Delta)}^{-1}\left\{\Big(\mathbf{V}\big(\Delta(v),\varepsilon\big)\Big)^{\!\frac{3}{4}},\,h_{l_r(\Delta)}\right\}
\!\bigg)\times\\
&\times
\partial_s\phi(v')
\,\text{tr}\bigg(\!
\tau^lh_{l_t(\Delta\!')}^{-1}\left\{\Big(\mathbf{V}\big(\Delta'(v'),\varepsilon\big)\Big)^{\!\frac{3}{4}},\,h_{l_t(\Delta\!')}\right\}
\!\bigg)\,\text{tr}\bigg(\!
\tau^mh_{l_u(\Delta\!')}^{-1}\left\{\Big(\mathbf{V}\big(\Delta'(v'),\varepsilon\big)\Big)^{\!\frac{3}{4}},\,h_{l_u(\Delta\!')}\right\}
\!\bigg).\!\!\!\!\!\!\!\!\!\!\!\!\!\!
\end{split}
\end{equation}

If the scalar field is sufficiently smooth one can write 
\begin{equation}
\partial_p\phi(v)
\approx\frac{1}{\varepsilon}\,\frac{e^{\phi_{v+\vec{e}_p}-\phi_{v}}-e^{\phi_{v}-\phi_{v-\vec{e}_p}}\!}2\,,
\end{equation}
where $\phi_{v+\vec{e}_p}$ is the field in the point $v+\vec{e}_p$, which is the nearest node of $v$ along the link $e_p$ of length $\epsilon$. Finally, applying it to the formula (\ref{der}) one gets 
\begin{equation}\label{der1bis}
\begin{split}
H^{(\phi)}_{der}=
&\,\frac{2^{13}}{3^4\lambda\,(\gamma\kappa)^4}\lim_{\varepsilon\to0}
\!\!\sum_{v\in\mathbf{V}(\Gamma)}\!\!\!N(v)\!\!\!\!\!\!\sum_{\Delta(v)=\Delta'(v)=v}\!\!\!\!\!\!\epsilon_{ijk}\epsilon_{pqr}\epsilon^i_{\,lm}\epsilon_{stu}
\times\\
&\times\frac{e^{\phi_{v+\vec{e}_p}-\phi_{v}}-e^{\phi_{v}-\phi_{v-\vec{e}_p}}\!}2\times \frac{e^{\phi_{v+\vec{e}_s}-\phi_{v}}-e^{\phi_{v}-\phi_{v-\vec{e}_s}}\!}2\times\\
&\times\text{tr}\Big(\!
\tau^jh_{l_q(\Delta)}^{-1}\left\{\big(\mathbf{V}\big(\Delta(v),\varepsilon\big)\big)^{\!\frac{3}{4}},\,h_{l_q(\Delta)}\right\}
\!\Big)
\,\text{tr}\Big(\!
\tau^kh_{l_r(\Delta)}^{-1}\left\{\big(\mathbf{V}\big(\Delta(v),\varepsilon\big)\big)^{\!\frac{3}{4}},\,h_{l_r(\Delta)}\right\}
\!\Big)\times\\
&\times\text{tr}\Big(\!
\tau^lh_{l_t(\Delta\!')}^{-1}\left\{\big(\mathbf{V}\big(\Delta(v),\varepsilon\big)\big)^{\!\frac{3}{4}},\,h_{l_t(\Delta\!')}\right\}
\!\Big)
\,\text{tr}\Big(\!
\tau^{m\!}h_{l_u(\Delta\!')}^{-1}\left\{\big(\mathbf{V}\big(\Delta(v),\varepsilon\big)\big)^{\!\frac{3}{4}},\,h_{l_u(\Delta\!')}\right\}
\!\Big).\!\!\!
\end{split}
\end{equation}


\section{Quantization of the scalar Hamiltonian constraint}\label{V}

The field contribution to the scalar constraint is quantized by the canonical procedure: the cubulation of the spatial manifold is given by the graph $\Gamma$ at which the state is based (links and nodes of the cubulations are links and nodes of $\Gamma$), while holonomies, volumes and matter variables are changed into quantum operators that act on states (\ref{state}) belonging to $\mathcal{H}_{kin}^{(tot)}$:
\begin{equation}\label{HCO1}
\hat{H}\!\ket{\Gamma;m_l,i_v;U_{\psi}}_{\!R}
=\big(\hat{H}^{(\phi)}_{kin}+\hat{H}^{(\phi)}_{der}+\hat{H}^{(\phi)}_{pot}\big)\!\ket{\Gamma;m_l,i_v;U_{\psi}}_{\!R}.
\end{equation}
The Poisson brackets in \eqref{kin2} and \eqref{der1bis} are replaced by commutators and the following condition is going to be used 
\begin{equation}
\text{tr}\bigg(\!
\tau^i\hat{h}_{l_p}^{-1}\left[\hat{\mathbf{V}}^n(R),\,\hat{h}_{l_p}\right]
\!\bigg)=
\text{tr}\bigg(\!
\tau^i\hat{h}_{l_p}^{-1}\hat{\mathbf{V}}^n(R)\,\hat{h}_{l_p}
\!\bigg)\,.
\end{equation}

The quantum operator corresponding to the kinetic part (\ref{kin2}) acts as follows:
\begin{equation}\label{kin2bis}
\begin{split}
\hat{H}^{(\phi)}_{kin}\!\ket{\Gamma;m_l,i_v;U_{\psi}}_{\!R}=
&-\frac{2^{21}\lambda}{3^2(16\pi\gamma G\hbar)^6}\lim_{\varepsilon\to0}
\!\!\!\!\sum_{v,v'\in\mathbf{V}(\Gamma)}\!\!\!\!\!\!N(v)\hat{\Pi}(v)\hat{\Pi}(v')\,\chi_{\varepsilon}(v,v')
\sum_{\Delta(v)}\,\sum_{\Delta'(v')}\!\!\epsilon_{ijk}\epsilon_{pqr}\epsilon_{lmn}\epsilon_{stu}\times\\
&\times\,
\text{tr}\bigg(\!
\tau^j\hat{h}_{l_q(\Delta)}^{-1}\Big(\hat{\mathbf{V}}\big(\Delta(v),\varepsilon\big)\Big)^{\!\frac{1}{2}}\hat{h}_{l_q(\Delta)}
\!\bigg)\times\\
&\times\text{tr}\bigg(\!
\tau^k\hat{h}_{l_r(\Delta)}^{-1}\Big(\hat{\mathbf{V}}\big(\Delta(v),\varepsilon\big)\Big)^{\!\frac{1}{2}}\hat{h}_{l_r(\Delta)}
\!\bigg)\,\text{tr}\bigg(\!
\tau^l\hat{h}_{l_s(\Delta\!')}^{-1}\Big(\hat{\mathbf{V}}\big(v'\!(\Delta\!'),\varepsilon\big)\Big)^{\!\frac{1}{2}}\hat{h}_{l_s(\Delta\!')}
\!\bigg)\times\\
&\times\text{tr}\bigg(\!
\tau^m\hat{h}_{l_t(\Delta\!')}^{-1}\Big(\hat{\mathbf{V}}\big(v'\!(\Delta\!'),\varepsilon\big)\Big)^{\!\frac{1}{2}}\hat{h}_{l_t(\Delta\!')}
\!\bigg)\,
\text{tr}\bigg(\!
\tau^n\hat{h}_{l_u(\Delta\!')}^{-1}\Big(\hat{\mathbf{V}}\big(v'\!(\Delta\!'),\varepsilon\big)\Big)^{\!\frac{1}{2}}\hat{h}_{l_u(\Delta\!')}
\!\bigg)
\!\ket{\Gamma;m_l,i_v;U_{\psi}}_{\!R},\!\!
\end{split}
\end{equation}
where $\hat{\mathbf{V}}$ is the volume operator (\ref{volume}) and $\hat{\Pi}(w)$ is the momentum operator, whose action in Schr\"{o}dinger representation reads
\begin{equation}\label{quantummomentum}
\begin{split}
\hat{\Pi}(w)\ket{\Gamma;m_l,i_v;U_{\psi}}_{\!R}
=-i\hbar\frac{\partial}{\partial\phi(w)}\ket{\Gamma;m_l,i_v;U_{\psi}}_{\!R}.
\end{split}
\end{equation}

Then, taking the limit $\varepsilon\to0$, in the expression (\ref{kin2bis}), one reach the scale at which there remains only the single node $v\equiv v'$ and the dependency on the regulator $\varepsilon$ is removed:
\begin{equation}\label{kin3}
\begin{split}
\hat{H}^{(\phi)}_{kin}\!\ket{\Gamma;m_l,i_v;U_{\psi}}_{\!R}=&-\frac{2^{15}\lambda}{3^2(8\pi\gamma l_P^2)^6}
\!\!\sum_{v\in\mathbf{V}(\Gamma)}\!\!\!N(v)\hat{\Pi}^2(v)
\!\!\!\!\!\!\sum_{\Delta(v)=\Delta'(v)=v}\!\!\!\!\!\!\!\epsilon_{ijk}\epsilon_{pqr}\epsilon_{lmn}\epsilon_{stu}\times\\
&\times\text{tr}\Big(\!
\tau^i\hat{h}_{l_p(\Delta)}^{-1}\big(\hat{\mathbf{V}}(v)\big)^{\!\frac{1}{2}}\hat{h}_{l_p(\Delta)}
\!\Big)\,
\text{tr}\Big(\!
\tau^j\hat{h}_{l_q(\Delta)}^{-1}\big(\hat{\mathbf{V}}(v)\big)^{\!\frac{1}{2}}\hat{h}_{l_q(\Delta)}
\!\Big)\times\\
&\times\text{tr}\Big(\!
\tau^k\hat{h}_{l_r(\Delta)}^{-1}\big(\hat{\mathbf{V}}(v)\big)^{\!\frac{1}{2}}\hat{h}_{l_r(\Delta)}
\!\Big)\,
\text{tr}\Big(\!
\tau^l\hat{h}_{l_s(\Delta\!')}^{-1}\big(\hat{\mathbf{V}}(v)\big)^{\!\frac{1}{2}}\hat{h}_{l_s(\Delta\!')}
\!\Big)\times\\
&\times\text{tr}\Big(\!
\tau^m\hat{h}_{l_t(\Delta\!')}^{-1}\big(\hat{\mathbf{V}}(v)\big)^{\!\frac{1}{2}}\hat{h}_{l_t(\Delta\!')}
\!\Big)\,
\text{tr}\Big(\!
\tau^n\hat{h}_{l_u(\Delta\!')}^{-1}\big(\hat{\mathbf{V}}(v)\big)^{\!\frac{1}{2}}\hat{h}_{l_u(\Delta\!')}
\!\Big)
\!\ket{\Gamma;m_l,i_v;U_{\psi}}_{\!R}.
\end{split}
\end{equation}

The result is a sum of subsystems, called basic cells, that extend over all nodes of the graph $\Gamma$. Each basic cell is a sum of elements acting on each node surrounded by six nearest neighbor nodes and it is labeled by the position of the central one. The example, with the central node $v_{x,y,z}$, is given by the illustration of state (\ref{state}). This cellular structure allows to restrict calculations to a basic cell and to give the final result as the sum over all cells.

The action of the operator \eqref{kin3} can be computed from the following expression (see appendix)
\begin{equation}\label{traceaction2}
\begin{split}
\text{tr}\big(\tau^i\hat{h}_{l_p}^{-1}\hat{\mathbf{V}}^n(v)\,\hat{h}_{l_p}\big)\!\ket{\Gamma;m_l,i_v;U_{\psi}}_{\!R}
=&\frac{i}{4}\left(8\pi\gamma l_P^2\right)^{\!\frac{3}{2}n}
\Big(\Sigma^{(q)}_{v}\Sigma^{(r)}_{v}\Big)^{\!\frac{n}{2}}
\Delta^{(p),\frac{n}{2}}_{v}\,\delta^{ip}
\ket{\Gamma;m_l,i_v;U_{\psi}}_{\!R},
\end{split}
\end{equation}
where we introduced 
\begin{equation}
\begin{split}
\Sigma^{(q)}_{v}&=\frac{1}{2}(j^{(q)}_{v}+j^{(q)}_{v-\vec{e}_q})\\
\Delta^{(p),n}_v&=\frac{1}{2^n}\left[\Big(\Big|j^{(p)}_v-1\Big|+j^{(p)}_{v-\vec{e}_p}\Big)^{n}-
\Big(\Big|j^{(p)}_v+1\Big|+j^{(p)}_{v-\vec{e}_p}\Big)^{n}\right]\,,
\end{split}
\end{equation}
$\vec{e}_{p}$ being the unit vector along the direction $p$, such that $j^{(p)}_{v-\vec{e}_p}$ is the spin number of the link along $p$ ending in $v$. 

Hence, the kinetic operator (\ref{kin3}) reads:
\begin{equation}\label{kin4}
\begin{split}
\hat{H}^{(\phi)}_{kin}\!\ket{\Gamma;m_l,i_v;U_{\psi}}_{\!R}=&\,\frac{2^3\lambda}{3^2(8\pi\gamma l_P^2)^{\frac{3}{2}}}
\sum_{v}\!N_{v}\,\hat{\Pi}^2_{v}\,
\Sigma^{(x)}_v\, \Sigma^{(y)}_v\, \Sigma^{(z)}_v\times\\
&\times\sum_{\Delta\!(v)}\,\sum_{\{p,q,r\}\in\Delta(v)}\!\!\!\!
\Delta^{(p),\frac{1}{4}}_v\, \Delta^{(q),\frac{1}{4}}_v \Delta^{(r),\frac{1}{4}}_v\,
\sum_{\Delta'(v)}\,\sum_{\{s,t,u\}\in\Delta'(v)}\!\!\!\!
\Delta^{(s),\frac{1}{4}}_v\, \Delta^{(t),\frac{1}{4}}_v \Delta^{(u),\frac{1}{4}}_v 
\!\ket{\Gamma;m_l,i_v;U_{\psi}}_{\!R}.
\end{split}
\end{equation}
The summation $\sum_{\{p,q,r\}}$ comes from $\sum_{i,j,k,p,q,r}\epsilon_{ijk}\epsilon_{pqr} \delta_{ip}\delta_{jq}\delta_{kr}$ in \eqref{kin3} (with Kronecker $\delta$'s arising from \eqref{traceaction2}) and extends over the 6 permutations of links in a given triple. The summation $\sum_{\Delta\!(v)}$ is due to the 8 possible choices of triples with mutual orthogonal links, which span the 8 tetrahedra $\Delta$ surrounding the six-valent node. Since $\Delta^{(p),\frac{1}{4}}_v\, \Delta^{(q),\frac{1}{4}}_v \Delta^{(r),\frac{1}{4}}_v$ is invariant under permutations of links and choice of tetrahedra, the final expression is just $(6\times 8)$ that of a single triple. In the same way, we treat $\sum_{\Delta'(v)}\,\sum_{\{s,t,u\}\in\Delta'(v)}$. Finally, we get $(6\times 8)^2$ times the eigenvalue of the product of three trace operators acting on a single triple:
\begin{equation}\label{kin5}
\begin{split}
\!\!\hat{H}^{(\phi)}_{kin}\!\ket{\Gamma;m_l,i_v;U_{\psi}}_{\!R}=&\,
\frac{2^{11}\lambda}{(8\pi\gamma l_P^2)^{\frac{3}{2}}}
\!\sum_{v}\!N_{v}\,\hat{\Pi}^2_{v}\,\,
\Sigma^{(x)}_v \Sigma^{(y)}_v \Sigma^{(z)}_v\,\, 
\Big(\Delta^{(x),\frac{1}{4}}_v \Delta^{(y),\frac{1}{4}}_v \Delta^{(z),\frac{1}{4}}_v
\Big)^{\!2}
\!\ket{\Gamma;m_l,i_v;U_{\psi}}_{\!R}.
\end{split}
\end{equation}

Similarly, the quantization of the derivative part of the scalar constraint operator (\ref{der1bis}) gives:
\begin{equation}\label{der2}
\begin{split}
\hat{H}^{(\phi)}_{der}\!\ket{\Gamma;m_l,i_v;U_{\psi}}_{\!R}=
&\,\frac{2^{13}}{3^4\lambda\,(8\pi\gamma l_P^2)^4}
\!\!\sum_{v\in\mathbf{V}(\Gamma)}\!\!\!N(v)\!\!\!\!\!\!\sum_{\Delta(v)=\Delta'(v)=v}\!\!\!\!\!\!\epsilon_{ijk}\epsilon_{pqr}\epsilon^i_{\,lm}\epsilon_{stu}
\times\\
&\times\frac{e^{\hat{\phi}_{v+\vec{e}_p}-\hat{\phi}_{v}}-e^{\hat{\phi}_{v}-\hat{\phi}_{v-\vec{e}_p}}\!}2\times \frac{e^{\hat{\phi}_{v+\vec{e}_s}-\hat{\phi}_{v}}-e^{\hat{\phi}_{v}-\hat{\phi}_{v-\vec{e}_s}}\!}2\times\\
&\times\text{tr}\Big(\!
\tau^j\hat{h}_{l_q(\Delta)}^{-1}\big(\hat{\mathbf{V}}(v)\big)^{\!\frac{3}{4}}\hat{h}_{l_q(\Delta)}
\!\Big)
\,\text{tr}\Big(\!
\tau^k\hat{h}_{l_r(\Delta)}^{-1}\big(\hat{\mathbf{V}}(v)\big)^{\!\frac{3}{4}}\hat{h}_{l_r(\Delta)}
\!\Big)\times\\
&\times\text{tr}\Big(\!
\tau^l\hat{h}_{l_t(\Delta\!')}^{-1}\big(\hat{\mathbf{V}}(v)\big)^{\!\frac{3}{4}}\hat{h}_{l_t(\Delta\!')}
\!\Big)
\,\text{tr}\Big(\!
\tau^{m\!}\hat{h}_{l_u(\Delta\!')}^{-1}\big(\hat{\mathbf{V}}(v)\big)^{\!\frac{3}{4}}\hat{h}_{l_u(\Delta\!')}
\!\Big)
\!\ket{\Gamma;m_l,i_v;U_{\psi}}_{\!R}.\!\!\!
\end{split}
\end{equation}

Next, using the expression for the trace (\ref{traceaction2}) one obtains:
\begin{equation}\label{der2bis}
\begin{split}
\hat{H}^{(\phi)}_{der}\!\ket{\Gamma;m_l,i_v;U_{\psi}}_{\!R}=&
\,\frac{2^{5}(8\pi\gamma l_P^2)^{\frac{1}{2}}}{3^4\lambda}\!\!
\sum_{v\in\mathbf{V}(\Gamma)}\!\!\!N(v)\sum_{\Delta\!(v)}\,\sum_{\{p,q,r\}\in\Delta(v)}\,
\sum_{\Delta'(v)}\,\sum_{\{s,t,u\}\in\Delta'(v)}\!\!\epsilon_{ijk}\epsilon_{pqr}\epsilon^i_{\,lm}\epsilon_{stu}
\delta_{qj} \delta_{rk} \delta_{lt} \delta_{mu}\times\\
&\times\frac{e^{\hat{\phi}_{v+\vec{e}_p}-\hat{\phi}_{v}}-e^{\hat{\phi}_{v}-\hat{\phi}_{v-\vec{e}_p}}\!}2\times \frac{e^{\hat{\phi}_{v+\vec{e}_s}-\hat{\phi}_{v}}-e^{\hat{\phi}_{v}-\hat{\phi}_{v-\vec{e}_s}}\!}2\times\\
&\times\Big(\Sigma^{(p)}_v\Big)^{\!\frac{3}{4}}\,\Big(\Sigma^{(q)}_v\Sigma^{(r)}_v\Big)^{\!\frac{3}{8}}\,
\Delta^{(q),\frac{3}{8}}_v\,\Delta^{(r),\frac{3}{8}}_v\,
\Big(\Sigma^{(s)}_v\Big)^{\!\frac{3}{4}}\,\Big(\Sigma^{(t)}_v\Sigma^{(u)}_v\Big)^{\!\frac{3}{8}}\,
\Delta^{(t),\frac{3}{8}}_v\,\Delta^{(u),\frac{3}{8}}_v 
\!\ket{\Gamma;m_l,i_v;U_{\psi}}_{\!R}.
\end{split}
\end{equation}
Since the expression within the summations is invariant under the exchange of $q\longleftrightarrow r$ (and of $t\longleftrightarrow u$), we get
\begin{equation}\label{der3}
\begin{split}
\!\!\!\hat{H}^{(\phi)}_{der}\!\ket{\Gamma;m_l,i_v;U_{\psi}}_{\!R}=
&\,\frac{2^7(8\pi\gamma l_P^2)^{\frac{1}{2}}}{3^4\lambda}
\sum_{v}\!N_{v}\,\Big(\Sigma^{(x)}_v\Sigma^{(y)}_v \Sigma^{(z)}_v\Big)^{\!\frac{3}{4}}
\!\!\!\sum_{\Delta\!'(v),\Delta(v)}\!\delta_{ps}\times\\
&\times
\Bigg[
\frac{e^{\hat{\phi}_{v+\vec{e}_p}-\hat{\phi}_{v}}-e^{\hat{\phi}_{v}-\hat{\phi}_{v-\vec{e}_p}}\!}2\times \frac{e^{\hat{\phi}_{v+\vec{e}_s}-\hat{\phi}_{v}}-e^{\hat{\phi}_{v}-\hat{\phi}_{v-\vec{e}_s}}\!}2 \times\\
&\times \Big(\Sigma^{(p)}_v \Sigma^{(s)}_v\Big)^{\!\frac{3}{8}}\Big(\Delta^{(q),\frac{3}{8}}_v \Delta^{(r),\frac{3}{8}}_v\Big)^{\!2}
\Bigg]
\!\ket{\Gamma;m_l,i_v;U_{\psi}}_{\!R}\,,
\end{split}
\end{equation}
where $\delta_{ps}$ arises because the $\delta$'s in \eqref{der2bis} force $p=i$ and $s=i$.
Note that the number of terms involved in the summations differs from the one in the kinetic part of the Hamiltonian. The summation extends over eight tetrahedra in the both cases, $\Delta\!'$ and $\Delta$, giving $8^2$ terms (as in \eqref{kin4}). However, the Kronecker delta $\delta_{ps}$ identifies one edge of $\Delta$ ($l_p$) with one edge of $\Delta'$ ($l_s$). Therefore, the action of Hamiltonian extends over eight tetrahedra $\Delta\!'(v)$ and then over the four tetrahedra $\Delta(v)$, which share the link $l_p$. 

Finally, contracted  indexes $p$ and $s$ give the summation of over three directions $\{x,y,z\}$ and the equation (\ref{der3}) becomes:
\begin{equation}\label{der4}
\begin{split}
\hat{H}^{(\phi)}_{der}\!\ket{\Gamma;m_l,i_v;U_{\psi}}_{\!R}=
&\,\frac{2^{11}(8\pi\gamma l_P^2)^{\frac{1}{2}}}{3^4\lambda}
\!\sum_{v}\!N_{v}\,
\Big(\Sigma^{(x)}_v \Sigma^{(y)}_v \Sigma^{(z)}_v \Big)^{\!\frac{3}{4}}
\times\\
&\times
\Bigg[
\bigg(\frac{e^{\hat{\phi}_{v}-\hat{\phi}_{v-\vec{e}_x}}\!-e^{\hat{\phi}_{v+\vec{e}_x}-\hat{\phi}_{v}}}2\bigg)^{\!\!2}
\Big(\Sigma^{(x)}_v \Big)^{\!\frac{3}{4}}
\Big(\Delta^{(y),\frac{3}{8}}_v \Delta^{(z),\frac{3}{8}}_v\Big)^{\!2}
+
\\
&+
\bigg(\frac{e^{\hat{\phi}_{v}-\hat{\phi}_{v-\vec{e}_y}}\!-e^{\hat{\phi}_{v+\vec{e}_y}-\hat{\phi}_{v}}}2\bigg)^{\!\!2}
\Big(\Sigma^{(y)}_v \Big)^{\!\frac{3}{4}}
\Big(\Delta^{(z),\frac{3}{8}}_v \Delta^{(x),\frac{3}{8}}_v\Big)^{\!2}
+
\\
&+
\bigg(\frac{e^{\hat{\phi}_{v}-\hat{\phi}_{v-\vec{e}_z}}\!-e^{\hat{\phi}_{v+\vec{e}_z}-\hat{\phi}_{v}}}2\bigg)^{\!\!2}
\Big(\Sigma^{(z)}_v \Big)^{\!\frac{3}{4}}
\Big(\Delta^{(x),\frac{3}{8}}_v \Delta^{(y),\frac{3}{8}}_v\Big)^{\!2}
\Bigg]\!\ket{\Gamma;m_l,i_v;U_{\psi}}_{\!R}.
\end{split}
\end{equation}

The same result can be also obtained in other way, directly form the formula \eqref{der2}, by the following counting: there are six possibilities for attaching a first triple of spins to the links $\{l_p,l_q,l_r\}\subset\Delta(v)$ times two possibilities for the second triple of links $\{l_s,l_t,l_u\}\subset\Delta\!'(v)$, while the summation over triangularizations extends over eight tetrahedra $\Delta\!'(v)$ and over four tetrahedra $\Delta(v)$ (which have the link $l_s$ as one of their edges).

As for the potential part \eqref{potential}, quantization is straightforward, since the only geometric part is the volume $\mathbf{V}(v)$. Using (\ref{volume}) one gets the expression:
\begin{equation}
\begin{split}
\hat{H}^{(\phi)}_{pot}\!\ket{\Gamma;m_l,i_v;U_{\psi}}_{\!R}
=
&\,\frac{1}{2\lambda}\!\sum_{v\in\mathbf{V}(\Gamma)}\!\!\!N(v)\hat{V}\big(\phi(v)\big)\hat{\mathbf{V}}(v)
\!\ket{\Gamma;m_l,i_v;U_{\psi}}_{\!R}
=\\
=&\,\frac{\left(8\pi\gamma l_P^2\right)^{\!\frac{3}{2}}}{2\lambda}
\!\sum_{v}\!N_{v}\hat{V}\big(\phi_{v}\big)
\Big(\Sigma^{(x)}_v \Sigma^{(y)}_v \Sigma^{(z)}_v\Big)^{\!\frac{1}{2}}\!\ket{\Gamma;m_l,i_v;U_{\psi}}_{\!R},
\end{split}
\end{equation}
where the operator $\hat{V}(\phi)$ should be properly defined in the polymer representation.

Then the action of the total scalar constraint operator (\ref{HCO1}) reads:
\begin{equation}\label{HCO2}
\begin{split}
\hat{H}^{(\phi)}\!\ket{\Gamma;m_l,i_v;U_{\psi}}_{\!R}
=&
\sum_{v}\!N_{v}
\bigg(
\frac{2^{11}\lambda}{(8\pi\gamma l_P^2)^{\frac{3}{2}}}
\,\Sigma_{v}^{(x)}\,\Sigma_{v}^{(y)}\,\Sigma_{v}^{(z)}
\Big(
\Delta_{v}^{\!(x),\frac{1}{4}}\,\Delta_{v}^{\!(y),\frac{1}{4}}\,\Delta_{v}^{\!(z),\frac{1}{4}}
\Big)^{\!2}
\,\hat{\Pi}^2_{v}
+
\\
&+
\frac{2^{11}(8\pi\gamma l_P^2)^{\frac{1}{2}}}{3^4\lambda}
\Big(
\Sigma_{v}^{(x)}\,\Sigma_{v}^{(y)}\,\Sigma_{v}^{(z)}
\Big)^{\!\frac{3}{4}}
\times\\
&\times
\!\bigg[
\Big(\Sigma_{v}^{(x)}\Big)^{\!\frac{3}{4}}
\Big(\Delta_{v}^{\!(y),\frac{3}{8}}\Big)^{\!2}
\Big(\Delta_{v}^{\!(z),\frac{3}{8}}\Big)^{\!2}
\left(
\frac{e^{\hat{\phi}_{v}-\hat{\phi}_{v-\vec{e}_x}}\!-e^{\hat{\phi}_{v+\vec{e}_x}-\hat{\phi}_{v}}
}2\right)^{\!\!2}
+\\&+
\Big(\Sigma_{v}^{(y)}\Big)^{\!\frac{3}{4}}
\Big(\Delta_{v}^{\!(x),\frac{3}{8}}\Big)^{\!2}
\Big(\Delta_{v}^{\!(z),\frac{3}{8}}\Big)^{\!2}
\left(\frac{
e^{\hat{\phi}_{v}-\hat{\phi}_{v-\vec{e}_y}}\!-e^{\hat{\phi}_{v+\vec{e}_y}-\hat{\phi}_{v}}
}2\right)^{\!\!2}
+\\&+
\Big(\Sigma_{v}^{(z)}\Big)^{\!\frac{3}{4}}
\Big(\Delta_{v}^{\!(x),\frac{3}{8}}\Big)^{\!2}
\Big(\Delta_{v}^{\!(y),\frac{3}{8}}\Big)^{\!2}
\left(\frac{
e^{\hat{\phi}_{v}-\hat{\phi}_{v-\vec{e}_z}}\!-e^{\hat{\phi}_{v+\vec{e}_z}-\hat{\phi}_{v}}
}2\right)^{\!\!2}
\bigg]
+
\\
&+
\frac{\left(8\pi\gamma l_P^2\right)^{\!\frac{3}{2}}}{2\lambda}
\Big(
\Sigma_{v}^{(x)}\,\Sigma_{v}^{(y)}\,\Sigma_{v}^{(z)}
\Big)^{\!\frac{1}{2}}
\hat{V}\big(\phi_{v}\big)
\!\bigg)\!\ket{\Gamma;m_l,i_v;U_{\psi}}_{\!R}.
\end{split}
\end{equation}
The expression above gives the action of $H^{\phi}$ on a quantum level and it is the starting point for the analysis of the dynamics of the scalar field. It is worth noting that all the coefficients within \eqref{HCO2} are analytic.    

\subsection{Large j limit}

Let us now perform the large $j$ limit of the formula (\ref{HCO2}) and outline how the eigenvalue of the quantum Hamiltonian coincides with the classical expression \eqref{Hamiltonianconstraint} at the leading order. To calculate this limit one can consider the following expansion for $j\gg \frac{1}{2}$:
\begin{equation}
\Delta_{v}^{\!(p),n}=-n\big(\Sigma_{v}^{(p)}\big)^{n-1}\!+O\big(j^{n-3}\big)\,,
\end{equation}
and gets for the expectation value $h^{(\phi)}:=\bra{\Gamma;m_l,i_v;U_{\psi}}\hat{H}^{(\phi)}\!\ket{\Gamma;m_l,i_v;U_{\psi}}_{\!R}$:
\begin{equation}\label{HCOlarge_j}
\begin{split}
h^{(\phi)}
\approx&
\sum_{v}\!N_{v}
\Bigg\{
\frac{\lambda}{2}\,
\mathbf{V}_{v}^{-1}
\,\Pi^2_{v}
+
\frac{1}{2\lambda}
\mathbf{V}_{v}
\Bigg[
\frac{\Sigma_{v}^{(x)}}{8\pi\gamma l_P^2\,\Sigma_{v}^{(y)}\,\Sigma_{v}^{(z)}}
\langle \hat{\Delta}^2_x\phi_{v}\rangle
+
\frac{\Sigma_{v}^{(y)}}{8\pi\gamma l_P^2\,\Sigma_{v}^{(x)}\,\Sigma_{v}^{(z)}}
\langle \hat{\Delta}^2_y\phi_{v}\rangle
+\\&\hspace{11.5 mm}+
\frac{\Sigma_{v}^{(z)}}{8\pi\gamma l_P^2\,\Sigma_{v}^{(x)}\,\Sigma_{v}^{(y)}}
\langle \hat{\Delta}^2_z\phi_{v}\rangle
\,\Bigg]
+
\frac{1}{2\lambda}
\mathbf{V}_{v}\,
\langle \hat{V}(\phi_v)\rangle
\!\Bigg\},
\end{split}
\end{equation}
where we introduced the eigenvalues $\mathbf{V}_{v}$ and $\Pi_v$ of the volume operator $\hat{\mathbf{V}}$ and of the momentum operator $\hat{\Pi}_v$
\begin{equation}
\mathbf{V}_{v}:=\Big(\big(8\pi\gamma l_P^2\big)^{\!3}\,\Sigma_{v}^{(p)}\,\Sigma_{v}^{(q)}\,\Sigma_{v}^{(r)}\Big)^{\!\frac{1}{2}}\qquad \Pi_v=\hbar \psi_v\,, 
\end{equation}
and the expectation values $\langle\Delta^2_i \phi\rangle$ and $\langle \hat{V}(\phi)\rangle$, {\it i.e.} 
\begin{equation}
\langle \hat{\Delta}^2_p\phi_{v}\rangle= \bra{U_{\psi}} \left(\frac{
e^{\hat{\phi}_{v}-\hat{\phi}_{v-\vec{e}_p}}\!-e^{\hat{\phi}_{v+\vec{e}_p}-\hat{\phi}_{v}}
}2\right)^{\!\!2} \!\ket{U_{\psi}}\qquad \langle \hat{V}(\phi)\rangle=\bra{U_{\psi}}\hat{V}(\phi)\!\ket{U_{\psi}}\,.
\end{equation}

Next, using the definition of the characteristic function \eqref{characteristic}, one gets
\begin{equation}
\begin{split}
\!\!h^{(\phi)}
\approx&
\lim_{\varepsilon\to0}\sum_{v}\frac{1}{\varepsilon^3}\!\int\!\!d^3u\,\chi_{\varepsilon}(v,u)\,N(v)
\Bigg(
\frac{\lambda}{2\sqrt{q(v)}}
\,\frac{\Pi^2(v)}{\varepsilon^3}
+
\\
&+
\varepsilon\frac{\sqrt{q(v)}}{2\lambda}
\bigg[
\frac{p^{1}(v)}{p^{2}(v)\,p^{3}(v)}
\langle \hat{\Delta}^2_x\phi_{v}\rangle
\!+
\frac{p^{2}(v)}{p^{1}(v)\,p^{3}(v)}
\langle \hat{\Delta}^2_y\phi_{v}\rangle
\!+
\frac{p^{3}(v)}{p^{1}(v)\,p^{2}(v)}
\langle \hat{\Delta}^2_z\phi_{v}\rangle
\bigg]
\!+
\varepsilon^3\frac{\sqrt{q(v)}}{2\lambda}
\langle  \hat{V}\big(\phi_v\big)\rangle
\!\Bigg),
\end{split}
\end{equation}
where $p^{(i)}(u)$ denote gravitational momenta at the point $u$, which are related to spin-numbers by the following relation  
\begin{equation}
p^{i}(v)\,\varepsilon^2\!=8\pi\gamma l_P^2\Sigma^{(i)}_{v}\,, 
\end{equation}
and $q=|p^{1}p^{2}p^{3}|$ is metric determinant.  

Let us assume to construct a proper semiclassical state for the scalar field variables\footnote{The construction of semiclassical states for a quantum geometry in the presence of a scalar field is in preparation, along the lines of what has been done in vacuum \cite{Alesci:2014rra}.}, such that expectation values and eigenvalues become classical quantities. Hence, the semiclassical value $h^{(\phi)}_{cl}$ of $h^{(\phi)}$ becomes, in terms of the original un-smeared variables, 
\begin{equation}
\begin{split}
h^{(\phi)}_{cl}
\approx&
\lim_{\varepsilon\to0}\sum_{v}\int\!\!d^3u\,\chi_{\varepsilon}(v,u)\,N(v)
\Bigg(
\frac{\lambda}{2\sqrt{q(v)}}
\,\pi^2(v)
+
\\
&+
\frac{\sqrt{q(v)}}{2\lambda}
\bigg[
\frac{p^{1}(v)}{p^{2}(v)\,p^{3}(v)}
\big(\partial_x\phi(v)\big)^{\!2}
+
\frac{p^{2}(v)}{p^{1}(v)\,p^{3}(v)}
\big(\partial_y\phi(v)\big)^{\!2}
+
\frac{p^{3}(v)}{p^{1}(v)\,p^{2}(v)}
\big(\partial_z\phi(v)\big)^{\!2}
\bigg]
+
\frac{\sqrt{q(v)}}{2\lambda}
V\big(\phi(v)\big)
\!\Bigg),
\end{split}
\end{equation}
and in the limit $\varepsilon\rightarrow 0$ we have $v=u$ and $\sum_{v}\!\int\!\!d^3u\,\chi_{\varepsilon}(v,u)=\int d^3u$ so finding 
\begin{equation}\label{HCOclassical}
\begin{split}
\!\!h^{(\phi)}_{cl}\!\rightarrow&\!
\int\!\!d^3u\,N(u)
\bigg[\!
\frac{\lambda}{2\sqrt{q}}
\,\pi^2(u)
+
\frac{\sqrt{q}}{2\lambda}
\bigg(
q^{11}
\big(\partial_x\phi(u)\big)^{\!2}
+
q^{22}
\big(\partial_y\phi(u)\big)^{\!2}
+
q^{33}
\big(\partial_z\phi(u)\big)^{\!2}
\bigg)\!
+
\frac{\sqrt{q}}{2\lambda}
\,V\big(\phi(u)\big)
\!\bigg],\!
\end{split}
\end{equation}
where we introduced the inverse components of the metric tensor in terms of $p$'s 
\begin{equation}
q^{11}=\frac{p^{1}}{p^{2}p^{3}}\,\qquad q^{22}=\frac{p^{2}}{p^{3}p^{1}}\,,\qquad q^{33}=\frac{p^{3}}{p^{1}p^{2}}\,. 
\end{equation}

The expression above clearly coincides with the classical expression \eqref{Hamiltonianconstraint} with the metric in the diagonal gauge. 

\section{Conclusions}\label{VI}
We defined the action of the scalar field Hamiltonian in the diffeomorphisms invariant Hilbert space of QRLG, whose quantum states are based at cuboidal graphs with attached $U(1)$ group elements. Hence, we adapt the procedure defined in \cite{Thiemann:1997rt} to a cubulation of the spatial metric and to the reduced holonomies and fluxes proper of QRLG. The scalar field was described in terms of point-holonomies, which live only at nodes of the graph. The resulting action of the scalar field Hamiltonian has been regularized via standard tools of the full theory and its matrix elements are analytic. This is a key point of the formulation, which outlines how the dynamic analysis can be carried out analytically in QRLG not only in vacuum \cite{Alesci:2013xd}, but also in the presence of a scalar field. We expect that this is the case also for other matter fields, whose introduction in QRLG will be the subject of  future developments.   

We also checked how in the large $j$ limit, the Hamiltonian eigenvalues approach the classical Hamiltonian at the leading order. The next-to-the-leading order corrections are pure quantum corrections, which will be discussed in conjunction with their possible phenomenological implications.

In particular, there are two kinds of applications of QRLG in the presence of a scalar field. The first one is to regard the scalar field as a clock-like field, defining the evolution of geometric degrees of freedom on a quantum level. This is analogous to what has been done in LQC \cite{Bojowald:2011zzb,Ashtekar:2011ni}. However, the quantization procedure described here is not equivalent to the canonical one adopted in LQC, thus we expect nontrivial results from this analysis. 

The second application concerns regarding the scalar field as an actual matter component of the thermal bath and to analyze the corrections to the classical dynamics. This investigation could be relevant for those inflationary scenarios, which are based on a scalar inflaton, since it can predict modifications to both the background expansion of the Universe and the behavior of scalar perturbations.      

\appendix
\section{}

In this appendix we compute the action of the operator $\text{tr}\big(\tau^i\hat{h}_{l_p}^{-1}\hat{\mathbf{V}}^n(v)\,\hat{h}_{l_p}\big)$ and we infer \eqref{traceaction2}. Let us choose for simplicity $p=z$, we have
\begin{equation}
\text{tr}\big(\tau^i\hat{h}_{l_z}^{-1}\hat{\mathbf{V}}^n(v)\,\hat{h}_{l_z}\big)=-\sum_{abd}\,(\tau_i)_{ab}\, (\hat{h}^{-1}_{l_z})_{bd}\,V^n\,(\hat{h}_{l_z})_{da}
\end{equation}
with $a,b,d$ indexes in the fundamental representation and $\tau$ SU(2) basis elements; in the basis that diagonalizes $\tau_z$, the holonomies are diagonal:
\begin{equation}
(\hat{h}_{l_z})_{da}=e^{ia\theta} \delta_{da}.
\end{equation}
When we apply this object to a state, the volume acts after the insertion of the holonomy $\hat{h}_{l_z}$, so it gives a coefficient $[\Sigma^{(x)}\,\Sigma^{(y)}\,(\Sigma^{(z)}+a)]^{n/2}$ and we get 
\begin{align}
&-\left[\Sigma^{(x)}\,\Sigma^{(y)}\right]^{n/2}\sum_{abd=-1/2}^{1/2}\,(\tau_i)_{ab}\, (\hat{h}^{-1}_{l_z})_{bd}\,\left(\Sigma^{(z)}+a\right)^{n/2}\,(\hat{h}_{l_z})_{da}=\nonumber\\
&=-\left[\Sigma^{(x)}\,\Sigma^{(y)}\right]^{n/2}\sum_{abd=-1/2}^{1/2}\,(\tau_i)_{ab}\, e^{-id\theta} \delta_{bd}\,\left(\Sigma^{(z)}+a\right)^{n/2}\,e^{ia\theta} \delta_{da}.
\end{align}
Now using the $\delta$'s we get that $a=d=b$, such that the two exponentials disappear and
\begin{equation}
=-\left[\Sigma^{(x)}\,\Sigma^{(y)}\right]^{n/2}\sum_a (\tau_i)_{aa}\,\left(\Sigma^{(z)}+a\right)^{n/2}
\end{equation}
which, by considering that the only $\tau$ with nonvanishing diagonal components is $\tau_z$, becomes
\begin{equation}
=\frac{i}{2}\left[\Sigma^{(x)}\,\Sigma^{(y)}\right]^{n/2}\,\delta^{iz} \sum_a a\,\left(\Sigma^{(z)}+a\right)^{n/2}=-\frac{i}{4}\left[\Sigma^{(x)}\,\Sigma^{(y)}\right]^{n/2}\,\delta^{iz}\,\left[\left(\Sigma^{(z)}+1/2\right)^{n/2}-\left(\Sigma^{(z)}-1/2\right)^{n/2}\right].
\end{equation}
From the expression above equation \eqref{traceaction2} for $p=z$ follows. 
For $p=x,y$, $h_{l_p}$ is diagonal modulo the rotations, which can be moved to $\tau_i$. As a consequence, we get the same result but with the rotated $\tau$, which means that the only nonvanishing contribution is for $i=x,y$.


{\acknowledgments
The work of FC was supported by funds provided by the National Science Center under the agreement DEC12
2011/02/A/ST2/00294.
The work of E.A. was supported by the grant of Polish Narodowe Centrum Nauki nr 2011/02/A/ST2/00300.}


\newpage
\begin{thebibliography}{99}

\bibitem{Alesci:2012md} 
  E.~Alesci and F.~Cianfrani,
  Europhys.\ Lett.\  {\bf 104}, 10001 (2013)
  [arXiv:1210.4504 [gr-qc]].

\bibitem{Alesci:2013lea} 
  E.~Alesci and  F.~Cianfrani,
  arXiv:1303.0762 [gr-qc].

\bibitem{Alesci:2013xd} 
  E.~Alesci and F.~Cianfrani,
  Phys.\ Rev.\ D {\bf 87}, no. 8, 083521 (2013)
  [arXiv:1301.2245 [gr-qc]].

\bibitem{Alesci:2015jca} 
  E.~Alesci and F.~Cianfrani,
  arXiv:1506.07484 [gr-qc].

\bibitem{Alesci:2013xya} 
  E.~Alesci, F.~Cianfrani and C.~Rovelli,
  Phys.\ Rev.\ D {\bf 88}, 104001 (2013)
  [arXiv:1309.6304 [gr-qc]].

\bibitem{Alesci:2014uha}
  E.~Alesci and F.~Cianfrani,
  Phys.\ Rev.\ D {\bf 90}, no. 2, 024006 (2014)
  [arXiv:1402.3155 [gr-qc]].

\bibitem{Alesci:2014rra}
  E.~Alesci and F.~Cianfrani,
  arXiv:1410.4788 [gr-qc].

\bibitem{Alesci:2015nja} 
  E.~Alesci and F.~Cianfrani,
  arXiv:1506.07835 [gr-qc].

\bibitem{Ashtekar:2004eh} 
  A.~Ashtekar and J.~Lewandowski,
  Class.\ Quant.\ Grav.\  {\bf 21}, R53 (2004)
  [gr-qc/0404018].

\bibitem{Rovelli:2004tv} 
  C.~Rovelli,
  Cambridge, UK: Univ. Pr. (2004) 455 p

\bibitem{Thiemann:2007zz} 
  T.~Thiemann,
  Cambridge, UK: Cambridge Univ. Pr. (2007) 819 p
  [gr-qc/0110034].

\bibitem{Bojowald:2011zzb} 
  M.~Bojowald,
  Lect.\ Notes Phys.\  {\bf 835}, pp.1 (2011).

\bibitem{Ashtekar:2011ni} 
  A.~Ashtekar and P.~Singh,
  Class.\ Quant.\ Grav.\  {\bf 28}, 213001 (2011)
  [arXiv:1108.0893 [gr-qc]].  

\bibitem{Banerjee:2011qu} 
  K.~Banerjee, G.~Calcagni and M.~Martin-Benito,
  SIGMA {\bf 8}, 016 (2012)
  [arXiv:1109.6801 [gr-qc]].

\bibitem{Ashtekar:2006rx} 
  A.~Ashtekar, T.~Pawlowski and P.~Singh,
  Phys.\ Rev.\ Lett.\  {\bf 96}, 141301 (2006)
  [gr-qc/0602086].

\bibitem{Bojowald:2011hd} 
  M.~Bojowald, G.~Calcagni and S.~Tsujikawa,
  Phys.\ Rev.\ Lett.\  {\bf 107}, 211302 (2011)
  [arXiv:1101.5391 [astro-ph.CO]].

\bibitem{Agullo:2012sh} 
  I.~Agullo, A.~Ashtekar and W.~Nelson,
  Phys.\ Rev.\ Lett.\  {\bf 109}, 251301 (2012)
  [arXiv:1209.1609 [gr-qc]].

\bibitem{Calcagni:2012vw} 
  G.~Calcagni,
  Annalen Phys.\  {\bf 525}, no. 5, 323 (2013)
  [Erratum-ibid.\  {\bf 525}, no. 10-11, A165 (2013)]
  [arXiv:1209.0473 [gr-qc]].

\bibitem{Thiemann:1997rt} 
  T.~Thiemann,
  Class.\ Quant.\ Grav.\  {\bf 15}, 1281 (1998)
  [gr-qc/9705019].

\bibitem{Thiemann:1997rq} 
  T.~Thiemann,
  Class.\ Quant.\ Grav.\  {\bf 15}, 1487 (1998)
  [gr-qc/9705021].


\bibitem{Ashtekar:2002sn} 
  A.~Ashtekar, S.~Fairhurst and J.~L.~Willis,
  Class.\ Quant.\ Grav.\  {\bf 20}, 1031 (2003)
  [gr-qc/0207106].

\bibitem{Ashtekar:2002vh} 
  A.~Ashtekar, J.~Lewandowski and H.~Sahlmann,
  Class.\ Quant.\ Grav.\  {\bf 20}, L11 (2003)
  [gr-qc/0211012].

\bibitem{Kaminski:2005nc} 
  W.~Kaminski, J.~Lewandowski and M.~Bobienski,
  Class.\ Quant.\ Grav.\  {\bf 23}, 2761 (2006)
  [gr-qc/0508091].

\bibitem{Kaminski:2006ta} 
  W.~Kaminski, J.~Lewandowski and A.~Okolow,
  Class.\ Quant.\ Grav.\  {\bf 23}, 5547 (2006)
  [gr-qc/0604112].

\bibitem{Arnowitt:1962hi} 
  R.~L.~Arnowitt, S.~Deser and C.~W.~Misner,
  Gen.\ Rel.\ Grav.\  {\bf 40}, 1997 (2008)
  [gr-qc/0405109].

\bibitem{Ashtekar:1986yd} 
  A.~Ashtekar,
  Phys.\ Rev.\ Lett.\  {\bf 57}, 2244 (1986).

\bibitem{Ashtekar:1994mh} 
  A.~Ashtekar and J.~Lewandowski,
  J.\ Math.\ Phys.\  {\bf 36}, 2170 (1995)
  [gr-qc/9411046].

\bibitem{Thiemann:1996aw} 
  T.~Thiemann,
  Class.\ Quant.\ Grav.\  {\bf 15}, 839 (1998)
  [gr-qc/9606089].
  
\bibitem{Gaul:2000ba}
  M.~Gaul and C.~Rovelli,
  Class.\ Quant.\ Grav.\  {\bf 18}, 1593 (2001)
  [gr-qc/0011106].

\bibitem{covariant} C. Rovelli, F. Vidotto, \emph{Covariant Loop Quantum Gravity: An Elementary Introduction to Quantum Gravity and Spinfoam Theory}, Cambridge University Press, 2014.

\bibitem{Domagala:2012tq} 
  M.~Domagala, M.~Dziendzikowski and J.~Lewandowski,
  arXiv:1210.0849 [gr-qc].

\end{thebibliography}
\end{document}